\documentclass[fleqn,usenatbib]{rasti}

\usepackage{newtxtext,newtxmath}

\usepackage[T1]{fontenc}

\DeclareRobustCommand{\VAN}[3]{#2}
\let\VANthebibliography\thebibliography
\def\thebibliography{\DeclareRobustCommand{\VAN}[3]{##3}\VANthebibliography}

\usepackage{graphicx}	
\usepackage{amsmath}	
\usepackage{multirow}
\usepackage{subcaption}

\title[Monte Carlo conformal prediction for radio galaxy classification]{Monte Carlo conformal prediction for quantifying uncertainty in radio galaxy classification under ambiguous ground truth}

\author[A. Walls et al.]{
Alex Walls,$^{1,\dagger}$\thanks{E-mail: alexandra.walls@postgrad.manchester.ac.uk (AW)}
James Barry,$^{1, \dagger}$
Devina Mohan$^{1}$
and Anna M.~M.~Scaife$^{1,2}$
\\
$^{1}$Jodrell Bank Centre for Astrophysics, University of Manchester, Manchester M13 9PL, UK\\
$^{2}$Alan Turing Institute, Euston Road, London, NW1 2DB, UK
}

\date{Accepted XXX. Received YYY; in original form ZZZ}

\pubyear{\the\year{}}

\begin{document}
\label{firstpage}
\pagerange{\pageref{firstpage}--\pageref{lastpage}}
\maketitle

\begin{abstract}
Dramatically increasing data volumes are forcing astronomers to adopt automated methods for the identification and classification of astronomical objects. Although deep-learning models are often well-suited to this task, obtaining a measure of uncertainty on their predictions is challenging. Here we consider the suitability of Monte Carlo conformal prediction (MCCP) set size and confidence as measures of model uncertainty for the astronomical classification of radio galaxies. We demonstrate this approach using model predictions from a pre-trained radio galaxy foundation model, fine-tuned on a smaller set of labelled radio galaxies. We calibrate the MCCP by obtaining annotator-derived soft label distributions, i.e. probability distributions over classes instead of single class assignments, for each of these labelled radio galaxies and compare the resulting set sizes and confidence scores to predictive entropy measures for each galaxy obtained using a supervised Bayesian deep-learning model trained using Hamiltonian Monte Carlo (HMC). The comparison reveals only a weak correlation between the measures.
We suggest that this indicates that 
MCCP and predictive entropy capture fundamentally different aspects of the total uncertainty, and conclude that the applicability of MCCP in this context is severely limited by the practical considerations associated with obtaining soft label distributions for specialised classifications.
\end{abstract}




\section{Introduction}

\def\thefootnote{\dagger}\footnotetext{equal contribution}\def\thefootnote{\arabic{footnote}}

Radio galaxies are a subtype of galaxy distinguished by large-scale regions of synchrotron emission at radio wavelengths \citep{radio-galaxies}. Synchrotron radiation from radio galaxies is emitted by bipolar, relativistic jets - narrow, collimated outflows of magnetized plasma ejected by the active galactic nucleus (AGN) - and the broader lobes that form when those jets decelerate as a result of interactions with the ambient medium \citep{jets-and-lobes}. In many systems, only the approaching jet is visible in radio maps due to Doppler deboosting of the emission from the receding jet \citep{relativistic-beaming}. Supersonic jets drive strong bow shocks in the ambient medium, which compress the plasma and amplify its magnetic field due to the flux-freezing condition for ideal magnetohydrodynamics, resulting in extremely bright hot spots at the end of well-defined lobes \citep{FRII-lobes}. In contrast, subsonic jets slow gradually and undergo turbulent mixing with the ambient medium, resulting in diffuse lobes that diminish in brightness with distance from the galaxy's centre \citep{FRI-lobes}. Lobes have a typical span of 50 kpc to 1 Mpc from end to end \citep{extent-of-radio-galaxies}, meaning they usually extend far beyond the 10 to 30 kpc stellar component of their host galaxies \citep{elliptical-size, disk-size}.

The large-scale morphology of radio galaxies can provide insight into their physical properties. The surface brightness distribution is directly linked to the kinetic power of jets \citep{jet-power, jet-power-2}, which can be used to determine the mass accretion rate of the AGN, as well as the temperature of the accreted gas \citep{AGN-accretion}. Spectral index gradients within the lobes can be used to estimate the active lifetime of the AGN \citep{spectral-aging}. The motion of a radio galaxy through a dense intracluster medium (ICM) exerts ram pressure on its jets and bends them into narrow tails, the curvature of which encodes information about both the host galaxy's velocity and the density of the ICM \citep{jet-bending, jet-bending-2}. The strong link between radio galaxy morphology and its underlying physics makes morphological classification schemes extremely useful, as they enable astronomers to probe the physical properties of galaxies from radio maps alone and make statistical comparisons across large populations.

The most widely-used morphological classification scheme for radio galaxies is the Fanaroff-Riley (FR) scheme \citep{FR-explain}, which originally placed radio galaxies into one of two classes based on the ratio of the distance between their brightest regions and their total extent, $R_{\rm FR}$. Radio galaxies with $R_{\rm FR} < 0.5$ were classified as FRI, while those with $R_{\rm FR} \geq 0.5$ were classified as FRII. When devising the scheme, \cite{FR} observed that members of each class appear to share morphological traits; FRI galaxies typically have diffuse lobes that diminish in brightness with distance from the AGN, while FRII galaxies typically exhibit well-defined lobes with bright hot spots. Due to this observation, visual inspection of large-scale morphology became the primary method of determining the FR classification of radio galaxies, as different survey resolutions and sensitivities can lead to inconsistent measurements of $R_{\rm FR}$ \citep{FR-revisit}. 
Some radio galaxies display features typical of different FR classes on each lobe, and these are referred to as hybrids \citep{FR-hybrids}.



Due to improvements in the sensitivity and angular resolution of modern radio telescopes such as ASKAP \citep{ASKAP} and MeerKAT \citep{MeerKAT}, the volume of data generated by radio observatories has increased dramatically in recent years \citep{increasing-data-volume}. This has made manual classification of radio galaxies in new sky surveys highly impractical, highlighting the need for an automated approach. Machine learning (ML) techniques have been explored as a potential solution. Convolutional neural networks (CNNs), which excel at image recognition, are uniquely well-suited to this application \citep{CNN-layers}. 

CNNs have seen extensive use for classifying radio galaxies according to the FR scheme, with models generally achieving a high level of accuracy \citep{CNNs-for-radio-galaxy-classification,Aniyan2017ClassifyingNetwork,ndungu2023}. Moreover, classification has also been demonstrated as a high-performing downstream task for CNN and transformer-based radio galaxy foundation models, which use self-supervised pre-training to learn a foundational representation that can be fine-tuned against a small labelled dataset to provide classifications \citep[e.g.][]{RGZ-BYOL, lastufka2024}. 
However, modern CNNs tend to be poorly calibrated and are often overconfident in their predictions \citep{modern-CNNs-overconfident}. As such, the output of a CNN cannot be interpreted as the probability that it is correct, which makes obtaining a measure of uncertainty on model predictions challenging. For radio galaxy classification, having no reliable measure of model uncertainty makes it impossible to distinguish ambiguous morphologies from more typical cases. If model classifications are used in source catalogues without associated measures of uncertainty, misclassified sources could bias any derived population statistics. Furthermore, since out-of-distribution data points are typically misclassified with high confidence \citep{high-confidence-out-of-distribution}, it is likely that novel morphologies - those that do not fit into the FR classification scheme - would go unnoticed. It is therefore vital to find valid methods of quantifying model uncertainty if neural networks are used to automate radio galaxy classification.

\cite{BNN-uncertainty} demonstrated that Bayesian neural networks (BNNs) can be used to obtain a measure of model uncertainty on radio galaxy classifications. While regular CNNs learn deterministic weight values, BNNs learn a probability distribution for each weight, meaning they will not necessarily always make the same prediction for the same image \citep{BNNs-expensive}. By performing multiple forward passes per image, the entropy of the distribution of predictions can be calculated and used as an uncertainty measure. However, a serious disadvantage of this method is its computational cost: BNNs require significantly more computational resources to train compared to a standard CNN, see e.g. \cite{BNNs-expensive-2}, or Table~1 of \cite{devina-neurips}. In addition, this method requires hundreds of forward passes per input, dramatically increasing their prediction latency compared to standard CNNs, which require only one forward pass per input \citep{BNNs-expensive}. As a consequence of these factors, BNNs do not scale well with model and dataset size. This limits their suitability for radio galaxy classification, where large volumes of data are the primary motivation for developing automated methods. Furthermore, exact Bayesian inference for neural networks is intractable, therefore all BNNs perform only \emph{approximate} Bayesian inference. Moreover, there will be some degree of model misspecification due to over-parametrisation. Consequently, in the limit of finite data and imperfectly specified models, BNNs do not provide any frequentist coverage guarantees. Therefore measures of uncertainty derived from them can be unreliable if Bayesian credible intervals are miscalibrated under model or prior misspecification and/or limited data.

Conformal prediction (CP) has emerged as an alternative method for quantifying model uncertainty by constructing prediction intervals with finite-sample coverage guarantees. Instead of a single class prediction, CP produces a prediction set that includes every class that scored above a certain threshold, which is calculated using a set of calibration data that is withheld from the model during training. The threshold is calculated such that the probability that the prediction set contains the correct class is equal to a user-specified value, providing a statistical grounding to the output of the model \citep{conformal-prediction, intro_to_CP}. The size of the prediction set can then be used as an indicator of model uncertainty, with larger prediction sets corresponding to higher uncertainty \citep{cp-uncertainty}. The most significant advantage of this method over BNNs is that it is not computationally intensive. In addition, CP does not require a specific type of model architecture. This means that it can be applied after the model has been trained, and the model does not need to be designed with its use in mind, making it a much more flexible method.

Conformal prediction has recently found its way into astronomy and has been used to construct statistically rigorous prediction intervals for more standard statistical models and significance measurements, and also for the output of deep learning models. \citet{ashton2024calibrating} show how CP can be used to calibrate the pipelines for detecting Compact Binary Coalescence (CBC) signals in gravitational wave astronomy by applying it to a Mock Data Challenge of LIGO-Virgo data. The authors use traditional significance estimates 
as heuristics and calibrate them using CP to enable uncertainty estimation in signal detection 
from multiple independent pipelines with different sensitivities.
\citet{pezoa2024} show how conformal quantile regression \citep[CQR;][]{alpha-calibration} can be used to quantify uncertainty in deep learning based classification of synthetic $\textrm{H}\alpha$ spectral lines of massive stars using a fully-connected neural network with ten layers. \citet{yong_mnras_cp} have also used CQR to quantify uncertainty in virial black hole mass predictions from ML models.
\citet{jones2024redshift} use CP to calibrate the photometric redshift (photo-z) predictions from a variational inference based Bayesian CNN trained on galaxy images from the Hyper
Suprime-Cam (HSC) survey. They also demonstrate that photo-z uncertainties improve outlier detection compared to previous non-image based photo-z predictors. Most recently, \citet{singer2025} use the transductive CP approach \citep{vovk2013transductive} to construct prediction intervals for a standard measurement error model and apply it to exoplanet data. 

However, as demonstrated by \cite{monte-carlo-cp}, 
CP is not appropriate for applications where the ground truth, i.e. the true class, is ambiguous. Radio galaxy classification falls into this category: training datasets are labelled manually by human astronomers who may disagree on a classification, and who cannot obtain different views of the galaxies to verify their labels; moreover, classification schemes such as that developed by \cite{FR} are subjective, and represent a human interpretation of incompletely understood and complex underlying physics. Therefore, the concept of a ``true'' label in observational astronomy is perhaps not as clear cut as in many other applications of deep learning \citep[e.g.][]{tagsnotboxes}.

To overcome this kind of limitation in applications of conformal prediction, \cite{monte-carlo-cp} introduced a modified form of CP called \emph{Monte Carlo conformal prediction} (MCCP), and demonstrated its statistical validity with a model trained on diagnostic dermatology images (another case of inherently ambiguous data). 

While MCCP has been successfully applied to several different cases of image classification with inherent ambiguity, including detection of aortic stenosis in echocardiogram images \citep{aortic-stenosis} and zero-shot predictions of web images \citep{zero-shot-web-images}, it has not yet been applied to astronomical image classification. In this work, we evaluate MCCP as a method of quantifying model  uncertainty in FR classification of radio galaxies. In Section~\ref{sec:cp} we outline the methodology of conformal prediction, and Monte Carlo conformal prediction; in Section~\ref{sec:data} we describe the construction of a soft label distribution for the MiraBest radio galaxy dataset; in Section~\ref{sec:model} we describe the classification model used in this work, as well as the model training; in Section~\ref{Sec:results} we compare the uncertainty measures from Monte Carlo conformal prediction with those from Bayesian deep-learning; in Section~\ref{sec:discussion} we discuss the impact of potential biases in the human annotations and the interpretation of the different uncertainty measures, and in Section~\ref{sec:conclusions} we draw our conclusions.

\section{Conformal prediction}
\label{sec:cp}
Conformal prediction (CP) is a statistical framework that allows the quantification of uncertainty without making any distributional assumptions \citep{intro_to_CP}. In this work, we use conformal prediction to refer to \emph{split} conformal prediction (also called \emph{inductive} conformal prediction), which makes use of an additional data split - the calibration set - which is withheld from the model during training \citep{set-sizes}. It will be stated explicitly when other variants of conformal prediction are discussed.

The only assumption of CP is that the calibration set and future test points are exchangeable, meaning they are drawn from the same distribution and that their joint probability is invariant under permutations - in other words, the order in which calibration samples are drawn must not convey any additional information \citep{exchangeability}. Ordering is not a concern for image classification tasks, as each sample is drawn independently.


In the context of classification, where a model, $f$, is being used to classify a test point, $X_{\rm test}$, into one of $K$ different classes, it is typical to have $f: X \rightarrow \Delta^K$, where $\Delta^K$ is the $K$-simplex (e.g. softmax output). The predicted class label, $\hat{Y}$, is then taken to be the index of the largest element in the simplex. CP augments such a model by using the withheld calibration set, $(X_i, Y_i)$, to return a prediction set, $C(X_{\rm test}) \subseteq [K]$ for the test point. A non-conformity score (for example, the softmax subtracted from 1) is computed for each label $k \in [K]$, which is used to quantify how unusual it would be for $k$ to be assigned to $X_{\rm test}$ based on the distribution of the non-conformity scores assigned to the true labels in the calibration set. A $p$-value for each label, $p_{k}$, is calculated from the fraction of calibration points with scores greater than or equal to the score for that label. The label $k$ is included in the prediction set if $p_{k}(X_{\rm test}) > \alpha$, where $\alpha \in [0,1]$ is a user-defined error rate. Averaging over all possible test points and draws of the calibration set, the probability that the true label, $Y_{\rm test}$, is included in the prediction set is
\begin{equation}
\label{eq:coverage guarantee}
    1 - \alpha \leq \mathbb{P}(Y_{\rm test} \in C(X_{\rm test})) \leq 1 - \alpha + \frac{1}{n + 1},
\end{equation}
where $n$ is the size of the calibration set \citep{intro_to_CP}. This probability is referred to as the \emph{marginal coverage guarantee}, and the size of the prediction set, $|C(X_{\rm test})|$, can be interpreted as a proxy for the uncertainty in the model prediction, $\hat{Y}$, with larger prediction sets indicating a higher level of uncertainty. When $n$ is sufficiently large, the marginal coverage guarantee is approximately $1 - \alpha$.

The marginal coverage guarantee in \eqref{eq:coverage guarantee} applies on average over the entire population. In practice, only the \emph{empirical coverage} on a finite test set is observed. Because empirical coverage is computed over a finite sample, it generally fluctuates around the $1 - \alpha$ level. For a calibration set of size $n$ and an infinite test set, the distribution of empirical coverage is
\begin{equation}
    \label{eq:coverage-distribution}
    \mathbb{P}(Y_{\rm test} \in C(X_{\rm test}) \sim \textrm{Beta}(n + 1 - l, l),
\end{equation}
where $l = \lfloor(n + 1) \alpha \rfloor$ \citep{coverage-distribution}. For a test set of size $n_{\rm test}$, the empirical coverage is the average of $n_{\rm test}$ trials with probability of success drawn from \eqref{eq:coverage-distribution}, which is a beta-binomial distribution,
\begin{equation}
    \label{eq:empirical-coverage-distribution}
    \mathbb{P}(Y_{\rm test} \in C(X_{\rm test}) \sim \frac{1}{n_{\rm test}} \textrm{Binom} (n_{\rm test}, \mu),
\end{equation}
where $\mu \sim \textrm{Beta} (n + 1 - l, l)$ \citep{intro_to_CP}. The width of this distribution can be used to determine the uncertainty on the empirical coverage.


%
It is important to note that the expression in \eqref{eq:coverage guarantee} holds only for the entire label distribution. While exchangeability is important to ensure the validity of the coverage guarantee, it is not sufficient to ensure the efficiency of the prediction sets produced by the CP algorithm. Coverage over a specific label may exceed or fall short of the user-specified value, and because of this, classes that are under-represented in the data are often significantly under-covered since the threshold score is calculated primarily from majority-class samples \citep{class-conditional}. For the same reason, common classes tend to be over-covered, producing overly-conservative prediction sets. Per-class coverage guarantee can be restored through the use of the class-conditional variant of conformal prediction, which splits the calibration set up by true label and calculates a separate threshold score for each class \citep{class-conditional-2}. However, this method sacrifices the marginal coverage guarantee and results in significantly larger prediction set sizes for a calibration set of the same size. Due to the limited volume of labelled radio galaxy data, 
class-conditional conformal prediction was not used for the work presented in this paper.

It is assumed that the calibration set and the test sample are drawn i.i.d. from the joint distribution of inputs and labels,
\begin{equation}
    (X_i, Y_i) \sim \mathbb{P}(X) \otimes \mathbb{P}(Y|X),
\end{equation}
where $\mathbb{P}(Y|X)$ is the true posterior label distribution, and $\otimes$ denotes the product measure; however, for datasets that have been annotated by humans, it is more often the case that labels are provided as one-hot vectors (distributions in which the majority-vote label is assigned a probability of 1, and all other labels are assigned a probability of 0): $\mathbb{P}_{\rm vote} (Y=y | X=x)$. This means that the CP coverage guarantee corresponds to the voted label posterior, rather than the true posterior. 

For some applications this is not problematic. For example, in many popular benchmark datasets such as MNIST \citep{mnist1, mnist2} or CIFAR10 \citep{cifar, cifar10}, the examples are quite unambiguous, meaning that there is unlikely to be a significant level of disagreement between annotators, and the class labels are completely distinct, meaning that a true class can be determined uniquely. 

Conversely, in many astronomy contexts classification is more subjective and a majority-voted posterior will differ significantly from the true conditional distribution, which typically is unknowable \citep[e.g.][]{walmsley2020}. In such scenarios, it has been shown that using the distribution of annotator labels as an approximation to the true posterior will result in better calibration of the coverage guarantee \citep{monte-carlo-cp}. This approach can be undertaken using a sampling-based method, referred to as Monte Carlo CP (MCCP).

\subsection{Monte Carlo conformal prediction}
\label{mccp}

For a test point $X_{\rm test}$ with true label $Y_{\rm test}$, CP produces a prediction set $C(X_{\rm test})$ that is guaranteed to meet the condition
described by Equation~\ref{eq:coverage guarantee}, where the coverage, $1-\alpha$, is set by the user \citep{intro_to_CP}. However, in cases where the true label of $X_{\rm test}$ is ambiguous, the coverage guarantee no longer holds. This limitation can be overcome through the use of MCCP \citep{monte-carlo-cp}.

The key feature of MCCP that ensures its validity even for inherently ambiguous data is the construction of the calibration set; for each point in the withheld calibration data, a `true' label is sampled $m$ times from a label distribution. This results in a calibration set consisting of $m$ subsets of $n$ points, each with data $x_{i}$ and sampled `true' label $y_{i}$. The random sampling of `true' labels for the calibration set means that inherent ambiguity in the labels is factored into the decision to include a label in the prediction set.

This approach requires the labelled data to have soft labels, i.e. normalized scores corresponding to each class, rather than one-hot labels where the correct class has a score of 1, and all other classes have a score of 0. Soft labels are usually obtained by aggregating the labels applied by many annotators. The challenges associated with this requirement in the context of astronomical labelling are discussed further in Section~\ref{zooniverse}.

The calibration points are passed through the trained model. For each point, the model outputs a set of logits, $\mathbf{f} = (f_{1}, ..., f_{K})$, where $K$ is the number of potential classes. Logits are converted to normalized scores, $\mathbf{p} = (p_{1}, ... , p_{K})$, by applying the softmax function,
\begin{equation}
    \label{eq:softmax}
    p_{i} = \frac{e^{f_{i}}}{\sum^{K}_{j=1}e^{f_{j}}}.
\end{equation}
A set of user-defined non-conformity scores, $\mathbf{s} = (s_{1}, ... , s_{n})$, is then created for each of the $m$ subsets of the calibration set. The most common choice of non-conformity score for classification tasks, and the one used in this work, is given by
\begin{equation}
    \label{eq:non-conformity}
    s_{i} = 1 - p_{y_{i}},
\end{equation}
where $p_{y_{i}}$ is the softmax score associated with the true label. Examples of alternatives to this choice of conformity score were explored in \cite{malz2025}, where small differences in optimality were seen depending on context, but improvements relative to the score described by Equation~\ref{eq:non-conformity} were typically found to be marginal.

%
%
%
%
When a test point is passed through the model, the non-conformity score for each label, $s_{k}$ is calculated using its associated softmax score, as in \eqref{eq:non-conformity}. For each subset $j\in[m]$, a $p$-value is assigned to each label by calculating the fraction of calibration points with non-conformity scores greater than or equal to the non-conformity score for the label:
\begin{equation}
    p_{k}^{j}(X_{\rm test}) = \frac{1 + |\{i : s_{i}^{j} \geq s_{k}(X_{\rm test})\}|}{n + 1}.
\end{equation}
The $p$-values for each label are averaged over all $m$ subsets to obtain
\begin{equation}
    \label{eq:average-pvalues}
    \bar{p_{k}} = \frac{1}{m} \sum_{j=1}^{m} p_{k}^{j}.
\end{equation}
The prediction set is then constructed by including all labels with averaged $p$-values greater than the chosen $\alpha$:
\begin{equation}
    \label{eq:prediction-set}
    C(X_{\rm test}) = \{ y \in [K] : \bar{p_{k}} > \alpha \}.
\end{equation}
However, as \citet{monte-carlo-cp} point out, the coverage guarantee then becomes
\begin{equation}
\label{eq:worse-coverage-guarantee}
    \mathbb{P}(Y_{\rm test} \in C(X_{\rm test})) \geq 1 - 2\alpha,
\end{equation}
since $\bar{p_{k}}$ is the average of dependent $p$-values \citep{average-pvals-1, average-pvals-2}. \citet{monte-carlo-cp} demonstrate that the original coverage guarantee can be recovered by estimating the cumulative distribution function of the averaged $p$-values, but doing so requires an additional data split. Due to the limited size of our labelled dataset, this approach was not feasible.

To recover the $1 - \alpha$ coverage guarantee, we instead replaced the averaging step in \eqref{eq:average-pvalues} with max-$p$ aggregation; the maximum $p$-value in $p_{k}^{j}$ was selected and used in place of $\bar{p_{k}}$ in \eqref{eq:prediction-set} to determine the inclusion of label $k$ in the prediction set. This method is identical to using each of the $m$ calibration subsets to obtain a different prediction set using standard CP, then taking the union of those prediction sets to obtain the final prediction set \citep{union-aggregation}. As this does not involve averaging the $p$-values, the original coverage guarantee is recovered.

\subsection{Interpreting prediction set size}
\label{sec:pred_set_size}

Conformal prediction can be used to bound conditional entropy, which captures the underlying intrinsic uncertainty of the data generating process under the true labelling distribution. \citet{correia2024an}
link variable sized list decoding, which is used in the context of error correcting codes in communication theory, to the guarantees provided by conformal prediction and apply information theoretic inequalities for list decoding to derive three different upper bounds on the conditional entropy of the data generating process. These upper bounds can be made differentiable and used to train ML models directly. Conditional entropy directly affects the expected prediction set size. \citet{dhillon2024expected} derive theoretical bounds on the marginal and conditional expected prediction set size for split conformal prediction and analyse how the bounds depend on factors such as non-conformity score, number of calibration data points and the user-specified error rate. 

CP 
thus provides a 
statistical framework for interpreting the model's softmax outputs, allowing the size of prediction sets to be used as a measure of model uncertainty. In cases where non-conformity scores cluster near zero (corresponding to high softmax scores for the true classes in the calibration set), the threshold score will be small, and most test images will produce prediction sets containing only one label. A larger spread in non-conformity scores, either due to inherent label ambiguity or flaws in model generalisation, will increase the threshold score, resulting in more multi-label prediction sets \citep{set-sizes}. The coverage can be calibrated by choosing the smallest possible value of $\alpha$ that achieves a satisfactory level of coverage while maintaining an acceptable balance of prediction set sizes. In general, this means minimizing the average prediction set size,
\begin{equation}
    \label{eq:average-coverage}
    \overline{|C|} = \frac{1}{N} \sum_{i=1}^N |C(X_i)|,
\end{equation}
where $N$ is the number of test samples \citep{alpha-calibration}. This means that test samples with prediction set sizes greater than one - indicating high uncertainty - can be flagged for manual review, as most test samples can proceed through an automated pipeline, thus keeping the number of samples requiring manual review low enough to be practical.

 
In machine learning, total uncertainty is often decomposed into epistemic uncertainty (uncertainty stemming from the model's lack of knowledge, which can be reduced with more data) and aleatoric uncertainty (noise inherent to the data that cannot be eliminated with more data) \citep{uncertainty-decomposition, hullermeier2021aleatoric}. MCCP prediction set sizes are primarily expected to capture aleatoric uncertainty since the non-conformity score is calculated using softmax values which are first-order probabilities and cannot capture epistemic uncertainty. However, the connection between prediction set size and aleatoric uncertainty has not been formalised. \citet{hagos2025performance} examine the ability of prediction set sizes to quantify aleatoric uncertainty arising from inherent ambiguity in multiple datasets due to overlapping classes and find that there is a very weak to weak correlation with human annotations. While multiple datasets with differing degrees of overlap and eight different deep learning models are considered to generate prediction sets with three different CP algorithms, MCCP, which is designed to capture human annotator behaviour, is not considered in that work. We examine the correlation between prediction set size from MCCP and entropy of human annotator distribution for our dataset in Section \ref{sec:uq_results}.

However, the inability of MCCP to separate the contribution of each source to the overall uncertainty makes MCCP a broad-brush technique for uncertainty estimation. In cases where it is desirable to understand the contribution of each source to the total uncertainty, for example when attempting to introduce strategies to reduce overall uncertainty, this is a disadvantage.
Furthermore, since the maximum size of a prediction set is the total number of possible classes, it can only convey $K$ different levels of uncertainty. This makes it a much less granular method than BNNs, which produce continuous posterior distributions over classes \citep{BNNs-expensive}. As a result, the prediction set size may struggle to represent subtle differences in model confidence, especially for tasks involving few classes.

In addition to prediction set size, we also examine the conformal confidence as an uncertainty measure. For a given test point, the conformal confidence is the coverage at which its prediction set changes from a single label to two labels \citep{conformal-confidence}. Unlike the prediction set size, which is limited to $K$ discrete levels of uncertainty, the conformal confidence is a continuous measure, and is therefore capable of capturing more subtle variations in the model's uncertainty. However, since it depends only on the non-conformity score of the second label, it is insensitive to the margin between the top label and the second label. As a result, conformal confidence may fail to reflect differences in uncertainty between points where the model is highly confident in the top label, and points where the top label has a non-conformity score only slightly better than that of the second label, which can make it more difficult to interpret than the prediction set size. We examine the correlation between the conformal confidence from MCCP and the uncertainty measure obtained from a BNN trained on the same data in Section \ref{sec:uq_results}.

\section{Distribution of annotator labels: MiraBest}
\label{sec:data}

\subsection{MiraBest}

The MiraBest machine learning dataset \citep{MiraBest} consists of 1256 images of radio galaxies pre-processed for deep learning tasks. The dataset was constructed using the sample selection and classification described in \citet{Miraghaei-and-Best}, who made use of the parent galaxy sample from \citet{Best2012OnProperties}. Optical data from data release 7 of the Sloan Digital Sky Survey \citep[SDSS DR7;][]{abazajian2009seventh} was cross-matched with NRAO VLA Sky Survey  \citep[NVSS;][]{condon1998} and Faint Images of the Radio Sky at Twenty-Centimeters  \citep[FIRST;][]{VLA-FIRST} radio surveys. Parent galaxies were selected such that their radio counterparts had an active galactic nucleus (AGN) host rather than emission dominated by star formation. To enable classification of sources based on morphology, sources with multiple components in either of the radio catalogues were considered. 

The morphological classification was done by visual inspection at three levels: (i) The sources were first classified as FRI/FRII based on the original classification scheme of \citet{FR}. Additionally, 35 \emph{Hybrid} sources were identified as sources having FRI-like morphology on one side and FRII-like on the other \citep{gopalkrishna2000}. Of the 1329 extended sources inspected, 40 were determined to be unclassifiable. (ii) Each source was then flagged as `Confident' or `Uncertain' to represent the degree of belief in the human classification and, although this qualification was not extensively explained in the original paper, \cite{BNN-uncertainty} have shown that it is correlated with model posterior variance over the dataset. 
We note that these “uncertain” MiraBest sources are generally fainter in terms of their peak and total flux density, but are not significantly smaller compared to the confident sources.

\begin{table}

\centering
\caption{MiraBest class-wise composition by number of objects in each class. \label{tab:mb_classes}}
	\begin{tabular}{cccc}
	\hline
      & \textbf{FRI} & \textbf{FRII} & \textbf{Hybrid} \\
    \hline
      Confident & 397 & 436 & 19 \\
     Uncertain & 194 & 195 & 15\\
    \hline
	\end{tabular}
\end{table}

To ensure the integrity of the ML dataset, 73 objects from the original 1329 extended sources identified in the catalogue were not included: (i) 40 unclassifiable objects; (ii) 28 objects with extent greater than the chosen image size of $150\times150$ pixels; (iii) 4 objects which were found in overlapping regions of the FIRST survey; (iv) 1 object which was uniquely classified (FRII Confident Diffuse) and therefore could not be represented in both the training and test sets. The composition of the final dataset is shown in Table~\ref{tab:mb_classes}. 

\begin{figure*}
\centerline{\includegraphics[width=0.3\textwidth,trim={2cm 0cm 2cm 0cm},clip]{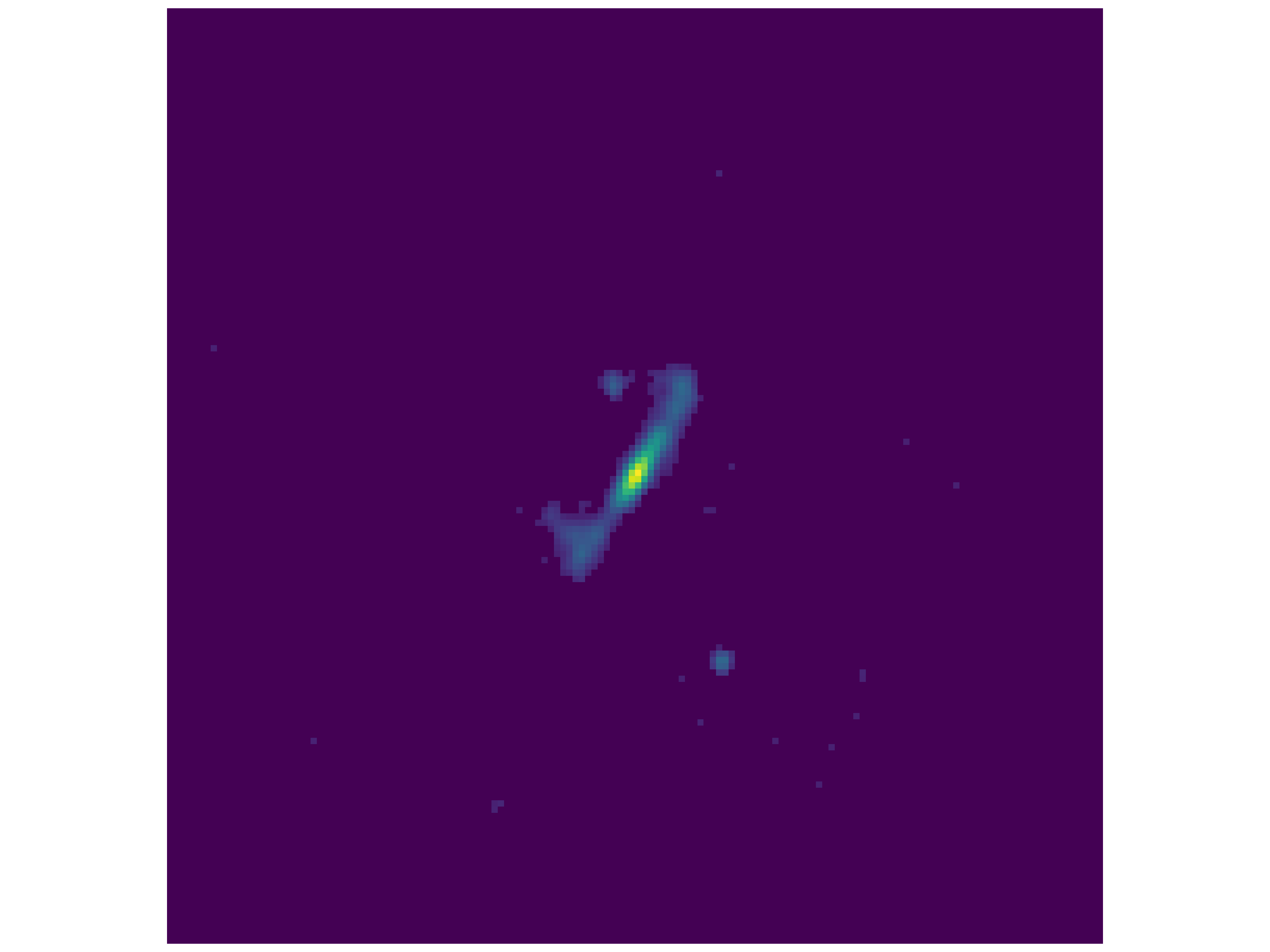}\qquad \includegraphics[width=0.3\textwidth,trim={2cm 0cm 2cm 0cm},clip]{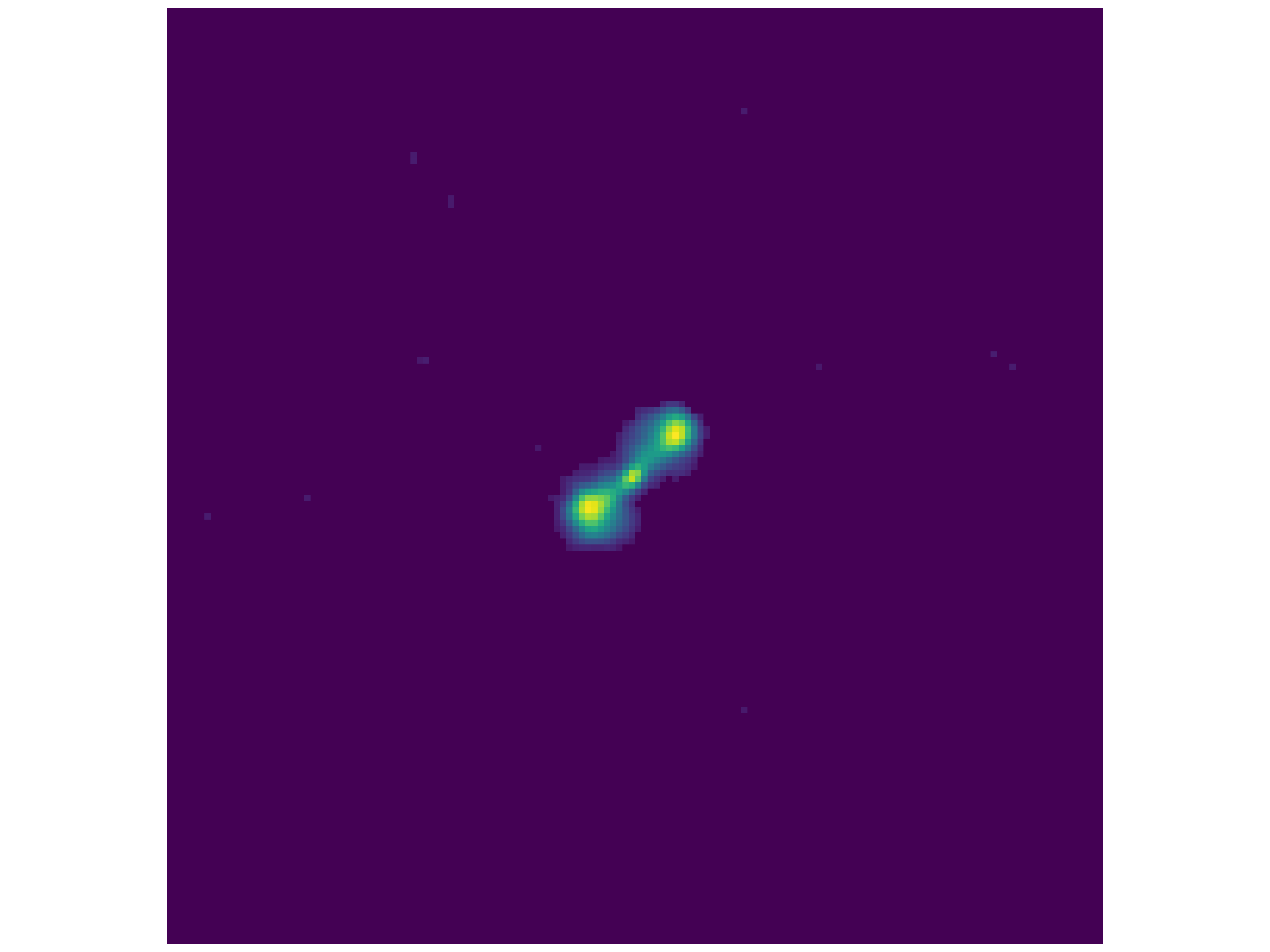}\qquad \includegraphics[width=0.3\textwidth,trim={2cm 0cm 2cm 0cm},clip]{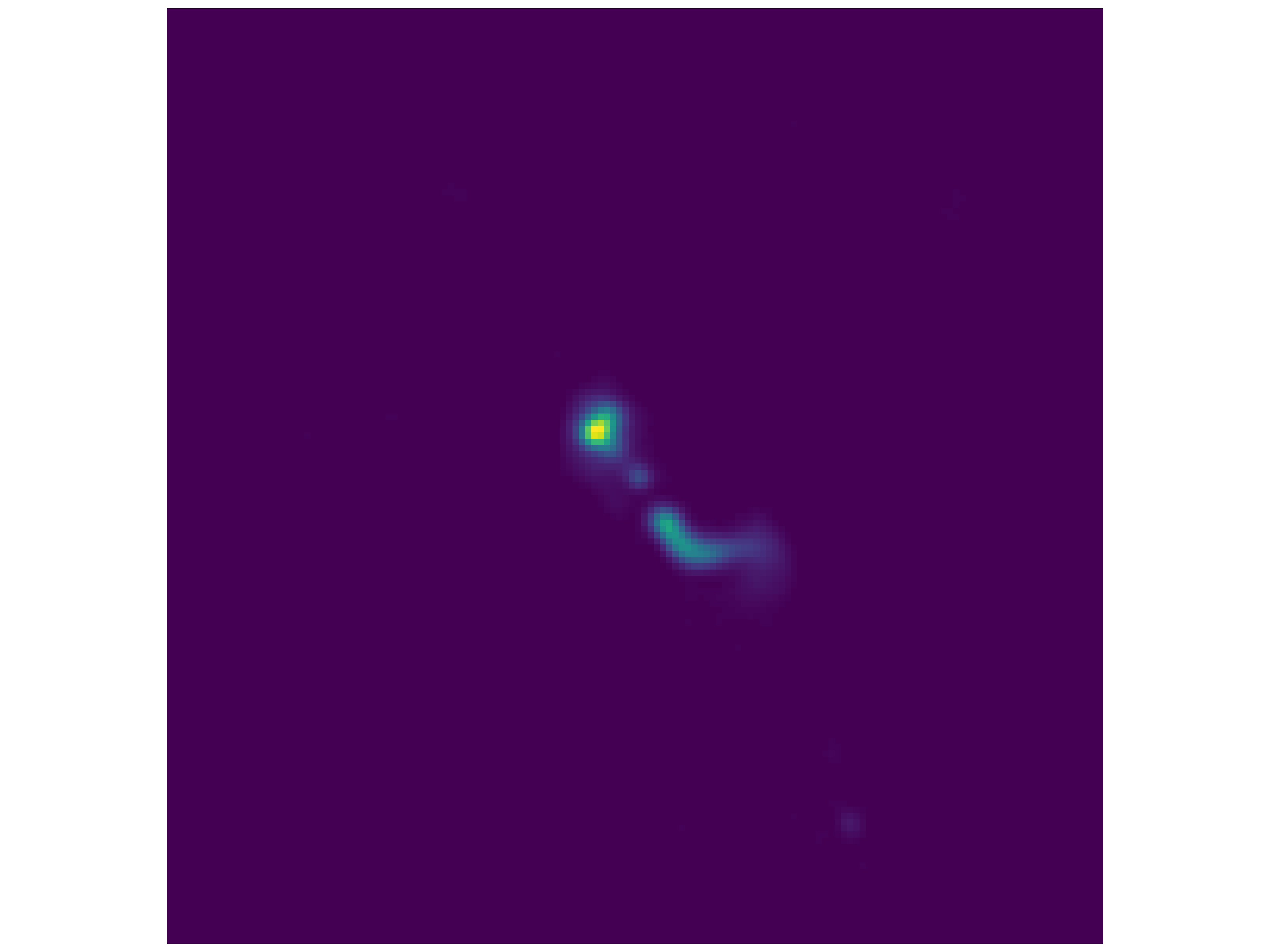}}
\centerline{(a) \hskip 0.3\textwidth (b) \hskip 0.3\textwidth (c)}
    \caption{\normalsize Radio images from the FIRST sky survey \citep{VLA-FIRST} depicting (a) an FRI galaxy, (b) an FRII galaxy, and (c) a hybrid galaxy. Each image has dimensions of 150$\mathrm{\times}$150 pixels, with each pixel corresponding to an angular size of 1.8 arcseconds.}
    \label{fig:MiraBest-examples}
\end{figure*}

Examples of radio galaxies from the MiraBest dataset belonging to each FR class, including a hybrid, are shown in Figure \ref{fig:MiraBest-examples}. Aside from its size relative to other labelled datasets, this dataset was chosen because it has previously been used to train CNNs to classify radio galaxies, with the resulting models achieving high levels of accuracy \citep{BNN-uncertainty, MiraBest-CNN-1}. In addition, it has been used to train BNNs to obtain uncertainty measures on FR classifications \citep{BNN}, which makes comparing the results of MCCP and BNNs straightforward.

However, as discussed in Section~\ref{mccp}, MCCP requires a calibration set with soft labels, while the MiraBest dataset has one-hot labels. Since there are currently no publicly available radio galaxy datasets with soft labels, it was necessary to source additional expert annotations for MiraBest in order to derive a soft label distribution. 

\subsection{Zooniverse survey}
\label{zooniverse}

\begin{figure}
    \centering
    \includegraphics[width=\linewidth]{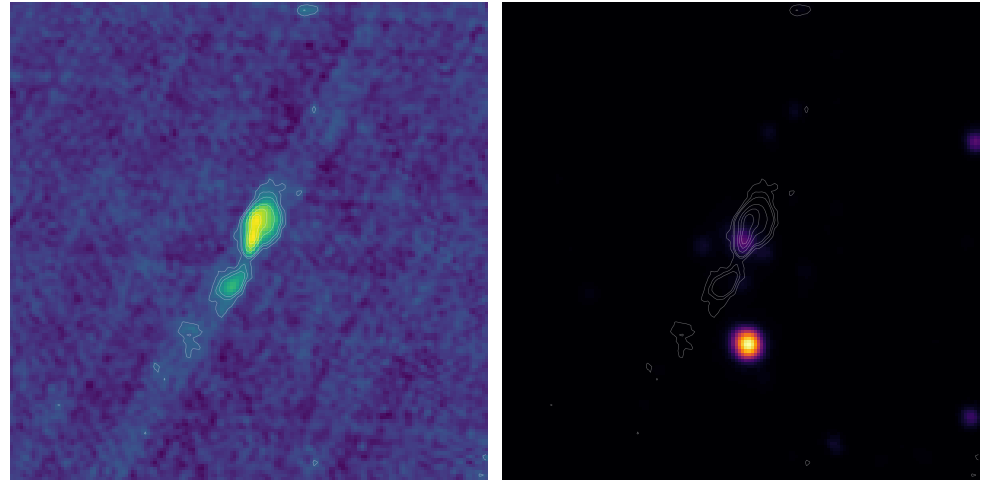}
    \caption{An example from the Zooniverse survey, with the FIRST image on the left and the WISE image on the right. Although both are greyscale images, they are both shown with a colour map to make their features more easily distinguishable.}
    \label{fig:zooniverse-example}
\end{figure}

To obtain a set of soft labels for the MiraBest dataset, we used the Zooniverse framework\footnote{\url{zooniverse.org}} to create a custom citizen science project from the 1256 FIRST images representing the MiraBest dataset. In this project, the FIRST images were displayed alongside infrared images of the same sky area from the Wide-field Infrared Survey Explorer (Band~1) \citep[WISE Band~1;][]{WISE}. Radio luminosity contours were displayed on the FIRST images at 3$\sigma$, 6$\sigma$, and 9$\sigma$, and then at 5 equally spaced levels between 9$\sigma$ and the maximum pixel value. The contours were also displayed on the WISE images. An example image is shown in Figure~\ref{fig:zooniverse-example}. The purpose of including the WISE images was to enable the annotators to identify the AGN host, which is bright at infrared wavelengths due to intense thermal radiation from hot dust around the accretion disk \citep{AGN-infrared}. This made it easier for annotators to determine whether a region of radio emission belonged to the radio galaxy itself. As is visible in Figure~\ref{fig:MiraBest-examples}(a), some FIRST images also contain radio-bright background objects and/or artefacts, and these can sometimes be confused with components of the radio galaxy itself. The corresponding WISE images provide vital context in these cases, especially where the morphology of the radio galaxy is more ambiguous.

Annotators were given the option to label each image as either an FRI galaxy, an FRII galaxy, or a hybrid. We chose to include the hybrid label despite the fact that previous work using the MiraBest dataset has focused only on FRI/FRII classification \citep{MiraBest-CNN-1}, as increasing the maximum prediction set size to 3 made the prediction set size a more fine-grained measure of uncertainty. If only FRI and FRII labels are included, only a single class prediction set (indicating minimum uncertainty) and a full prediction set (indicating maximum uncertainty) are possible. Including a third label increases the set size and allows for options between these extremes.

\begin{figure}
    \centering
    \includegraphics[width=\linewidth]{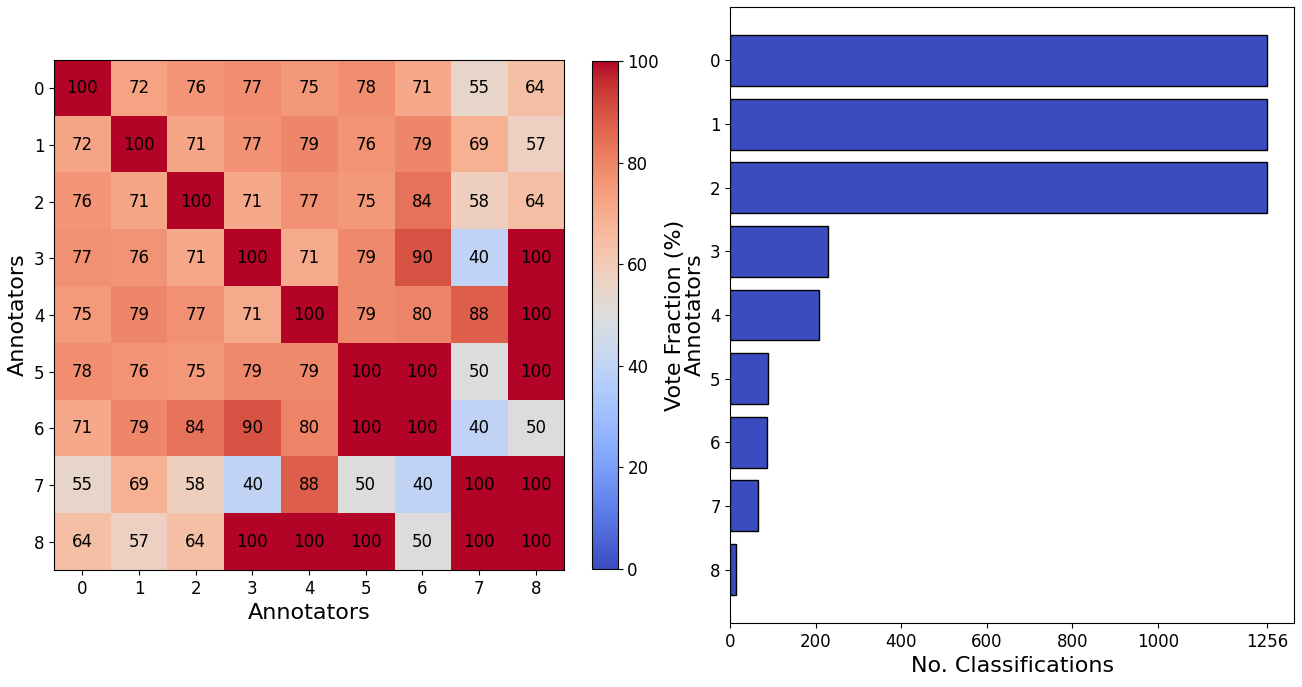}
    \caption{A heatmap showing the vote fraction between annotators as a percentage (left), and a plot showing the number of annotations provided by each annotator (right).}
    \label{fig:heatmap}
\end{figure}

The survey was sent to 14 radio astronomers with experience in classifying radio galaxies. A total of 4641 annotations were received from 8 annotators. The average number of annotations per image was 3.7. The original MiraBest labels were included as an additional set of annotations, increasing the total number of annotations to 5897 and the average number of annotations per image to 4.7. Although small, this number is consistent with the average annotations per image used in previous work utilising MCCP \citep[e.g.][]{monte-carlo-cp, aortic-stenosis}. The level of agreement between annotators and the number of annotations received per annotator are shown in Figure~\ref{fig:heatmap}. Since annotator 0 represents the original MiraBest labels, only two additional annotators labelled the entire dataset. Annotators generally had a high level of mutual agreement, with the exception of annotator 7. However, we note that annotator 7 provided only a small number of annotations. 


Several annotators noted that the low resolution of the FIRST images made visual galaxy classification challenging. Although higher resolution surveys existed at the time MiraBest was constructed, none combined the required sky coverage and sensitivity to construct a statistically robust sample of radio galaxies for large-scale morphology studies. Although the 5 arcsecond angular resolution of FIRST is sufficient to distinguish FRI and FRII features in larger sources, it can struggle to resolve fine morphological detail and the features of sources with smaller angular size. Multiple annotators stated that they defaulted to classifying sources as FRII in cases where the resolution made the morphology ambiguous. This is because at the largest distances, only the most luminous sources are detected, and FRII sources are typically more luminous than FRI sources \citep{FRII-luminous}. Additionally, the annotators noted that when source extent is small, a compact double structure makes an FRII classification more intuitively plausible in the absence of visible diffuse tails. The potential impact of these behaviours on the results is discussed further in Section~\ref{human-behaviour}.
\subsection{Data splits}
\label{data-splits}
In order to use MCCP, a subset of the data needed to be reserved as the calibration set and withheld from the model during training. A further withheld test set was also necessary in order to evaluate the model's performance on unseen data, which is important for verifying that the model will generalise to previously unseen data. We designated 12.5\% of the data as the calibration set, 12.5\% as the test set, and the remaining 75\% as the training set. Each split was stratified by class to preserve label proportions. We note that the training data split is slightly smaller than the 80\% typically considered ideal when using CP \citep{data-split}. This reduction was necessary to ensure that the calibration set and test set were representative of the entire dataset, as, while there are a roughly equal split of FRI and FRII galaxies, only 34 out of the total 1256 galaxies are hybrids.  Potential impacts of the smaller training set on the performance of the model are discussed further in Section~\ref{human-behaviour}.

The original MiraBest labels also contain `confident' and `uncertain' designations applied by \cite{Miraghaei-and-Best}. Due to this, the test set can be split further into a confident test set and an uncertain test set, and performance on each evaluated separately.

\section{Classification model}
\label{sec:model}

\subsection{Self-supervised learning}
\label{finetuning}
Although the MiraBest set is one of the largest labelled radio galaxy datasets, it is significantly smaller than typical datasets used in image recognition, which usually fall within the range of $10^{4}$-$10^{6}$ images \citep{dataset-size}. For this reason, we chose to use self-supervised learning (SSL), a form of representation learning where a model is initially pre-trained on unlabelled data. Instead of a fully-connected layer that maps the feature vector directly to classifications, SSL networks have two fully-connected layers that take the feature vector and map it to an embedding space (a projection head). The embedding space is a continuous, multidimensional space in which each image is represented by a single data point. The model learns by adjusting its embedding space such that augmentations of the same image are forced closer together, while different images are pushed farther apart. The result is an embedding space where images with similar features form tight clusters. The model can then be fine-tuned on a smaller labelled dataset, aligning its pre-learned representation with discrete classes \citep{SSL}.

For fine-tuning, the projection head is replaced with a classification head and the pre-trained model is trained further (fine-tuned) on labelled data. During fine-tuning, batches of labelled data are passed through the model to compute the loss, which is a scalar function that quantifies how well the model's predictions match the labels. 
Fine-tuned SSL models generally perform better than supervised learning models (where models learn directly from labelled data) in cases where the volume of labelled data is small. This is because the model is able to learn general features from a vast set of unlabelled data. When it is fine-tuned on the smaller set of labelled data, it already recognises patterns that are useful for distinguishing different classes, resulting in better performance than when it is trained from scratch on the same small set of labelled data \citep[see e.g.][]{SSL-better}. This effect is so significant that self-supervised CNNs can outperform supervised CNNs even in cases where the pre-training data is unrelated to the classification task, as the model still learns general-purpose, transferable representations of the data \citep{SSL-unrelated}. In this work we use the RGZ BYOL 
SSL radio galaxy model from \cite{RGZ-BYOL} as our pre-trained foundation.

\subsection{RGZ BYOL}

\cite{RGZ-BYOL} used the Bootstrap Your Own Latent (BYOL) \citep{BYOL} architecture to pre-train a self-supervised ResNet-18 model on the RGZ DR1 dataset \citep{RGZ-DR1}, which consists of approximately $10^{5}$ unlabelled radio galaxy images from the Australia Telescope Large Area Survey \citep[ATLAS;][]{Mao_2012} and the Faint Images of the Radio Sky at Twenty-Centimeters \citep[FIRST;][]{VLA-FIRST}. The model was then fine-tuned on the confidently-labelled FRI and FRII samples from the MiraBest dataset. When compared to a purely supervised CNN trained from scratch on the same data, the self-supervised model reduced the misclassification rate by 50\%. Furthermore, the self-supervised model achieved a high level of accuracy even when training on only a small subset of the MiraBest data.

In this work we adopted the pre-trained BYOL model from \cite{RGZ-BYOL} and fine-tuned it on the MiraBest data (including hybrids), using the data splits described in Section~\ref{data-splits}. The majority-voted classes from the Zooniverse survey were used as one-hot labels for this fine-tuning. Training was performed with a batch size of 64, and the AdamW optimiser \citep{Loshchilov2017DecoupledWD} with a weight decay of 5 $\times$ $10^{-3}$. The learning rate was initialized at 2.5 $\times$ $10^{-4}$ and a cosine decay schedule was applied, smoothly reducing the learning rate toward zero over 300 epochs, at which point the training was stopped. This smooth decay in learning rate meant that weight updates became smaller over time, helping the weights to settle without sudden jumps.

\section{Results}
\label{Sec:results}

\subsection{Model performance}
\begin{figure}
    \includegraphics[width=\linewidth,trim={0cm 0.3cm 0cm 0.1cm},clip]{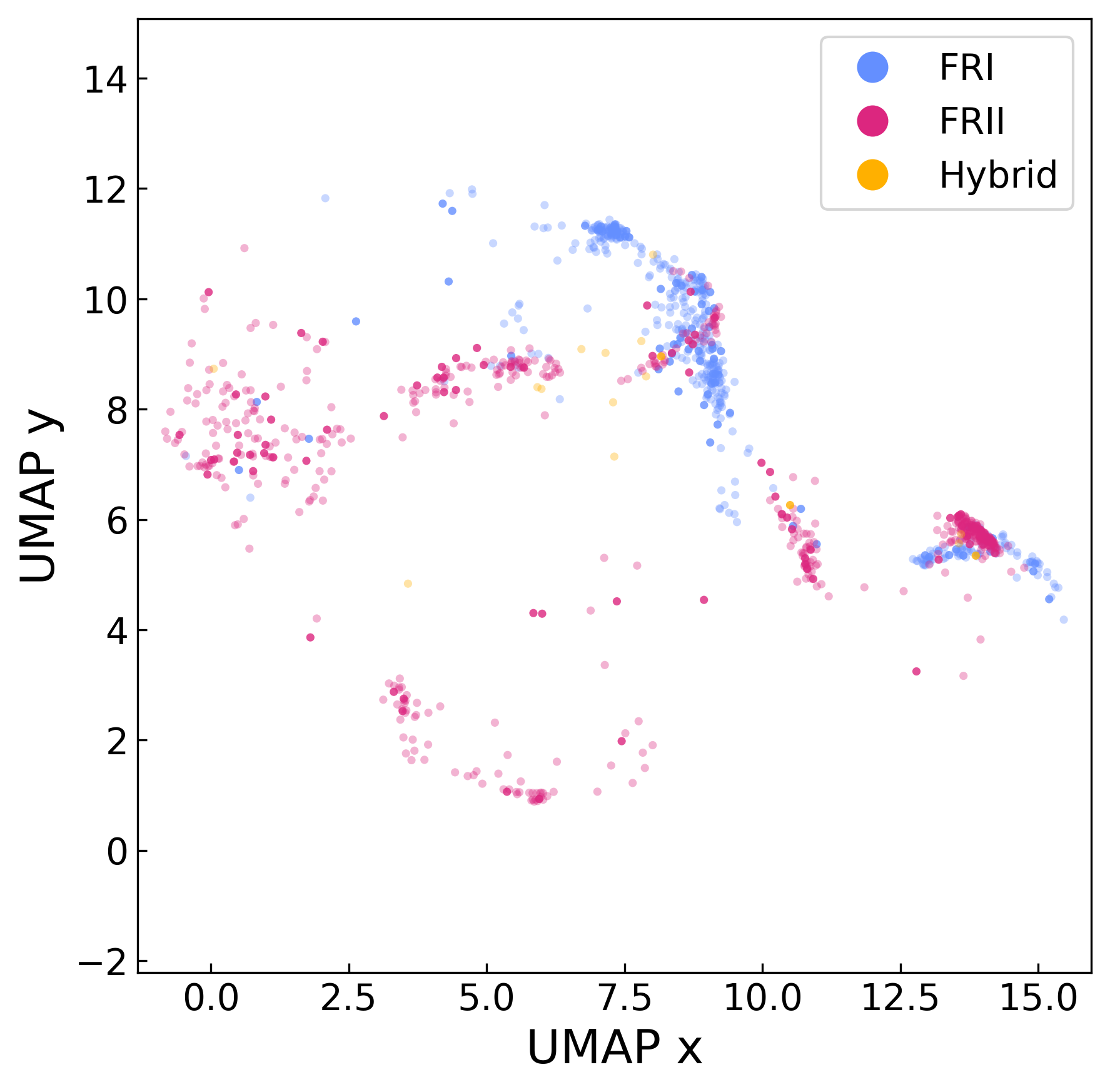}
    \centerline{(a)}
    \includegraphics[width=\linewidth,trim={0cm 0.3cm 0cm 0.1cm},clip]{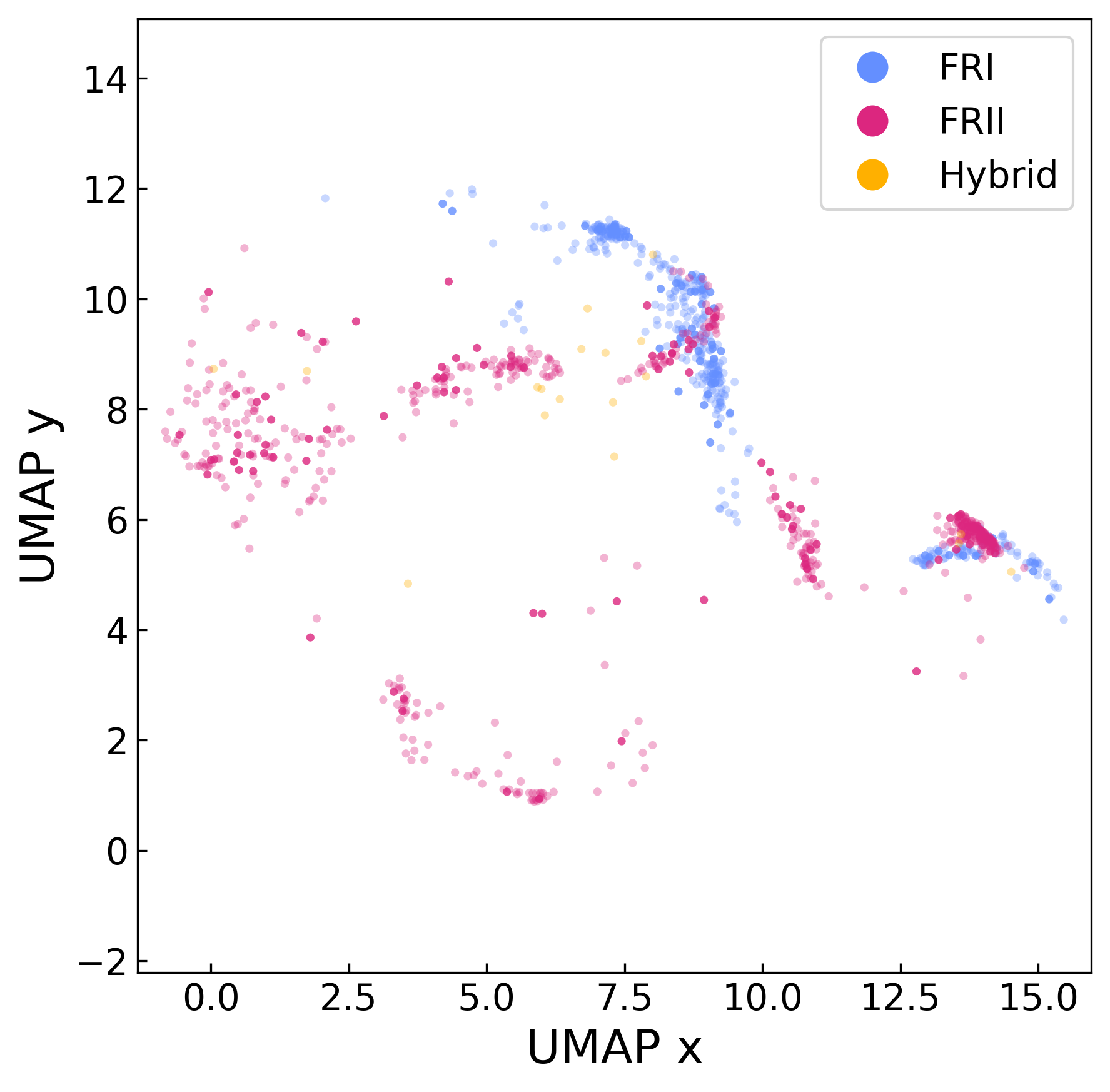}
    \centerline{(b)}
    
    \caption{Embedding of MiraBest data points from the model finetuned on majority-voted labels, colour-coded by (a) majority-vote annotator label and (b) model prediction. Points that were included in the training set are plotted with lower opacity. Note that the x and y dimensions here have no physical interpretation.}
    \label{fig:embedding-1}
\end{figure}

After fine-tuning, the model achieved 76\% accuracy on the uncertain test set and 92\% accuracy on the confident test set. To visualise the model's embedding space in two dimensions, we used UMAP (Uniform Manifold Approximation and Projection), a dimensionality reduction technique that preserves the spatial relationship between data points \citep{UMAP}. The embedding of the MiraBest data is shown in Figure~\ref{fig:embedding-1}, with data points colour-coded by majority-vote annotation (Figure~\ref{fig:embedding-1}(a)) and by model prediction (Figure~\ref{fig:embedding-1}(b)). The embeddings demonstrate that the model generally distinguishes the three classes well.

\begin{figure}
    \includegraphics[width=\linewidth,trim={0cm 0.3cm 0cm 0.1cm},clip]{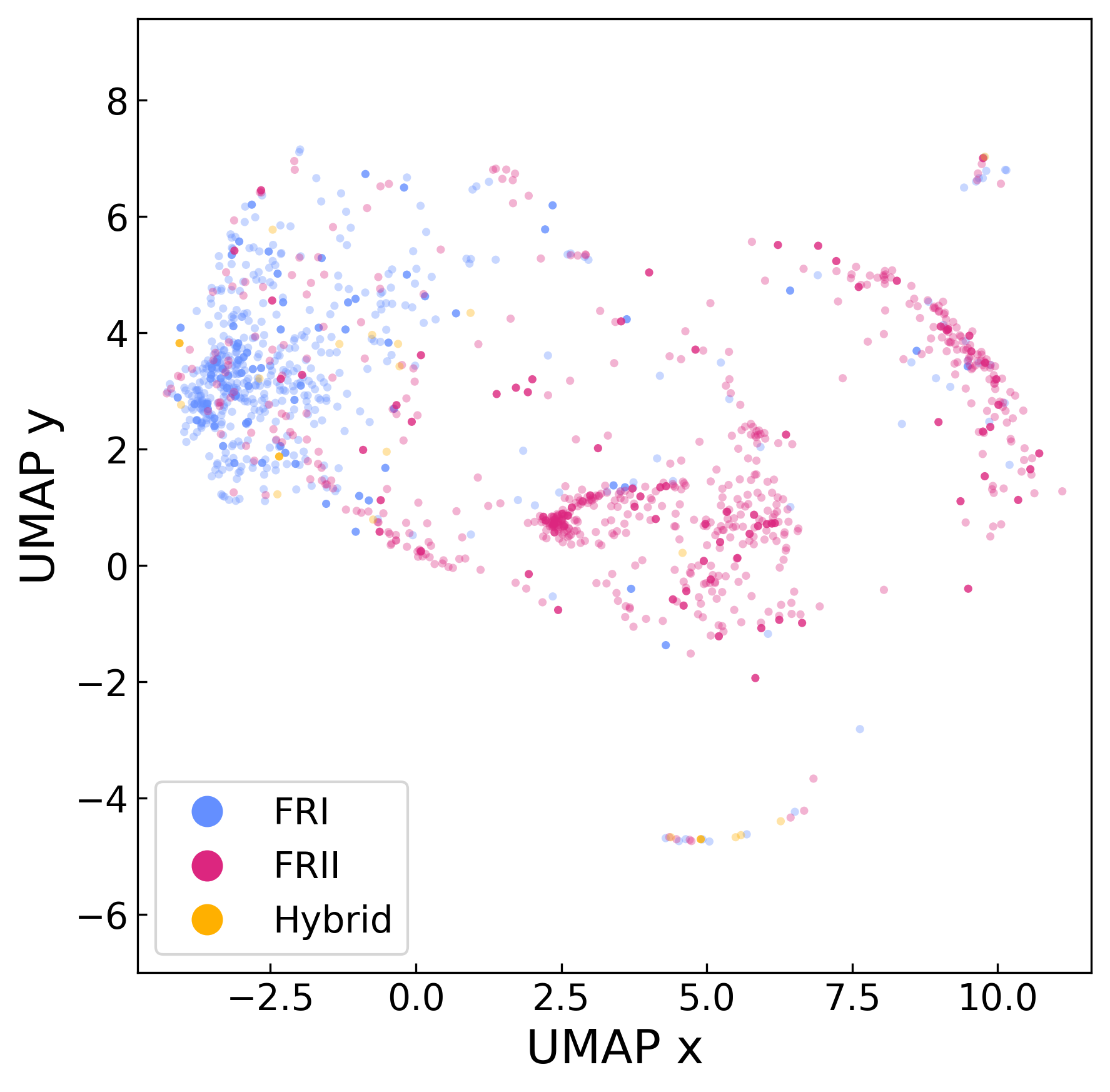}
    \centerline{(a)}
    \includegraphics[width=\linewidth,trim={0cm 0.3cm 0cm 0.1cm},clip]{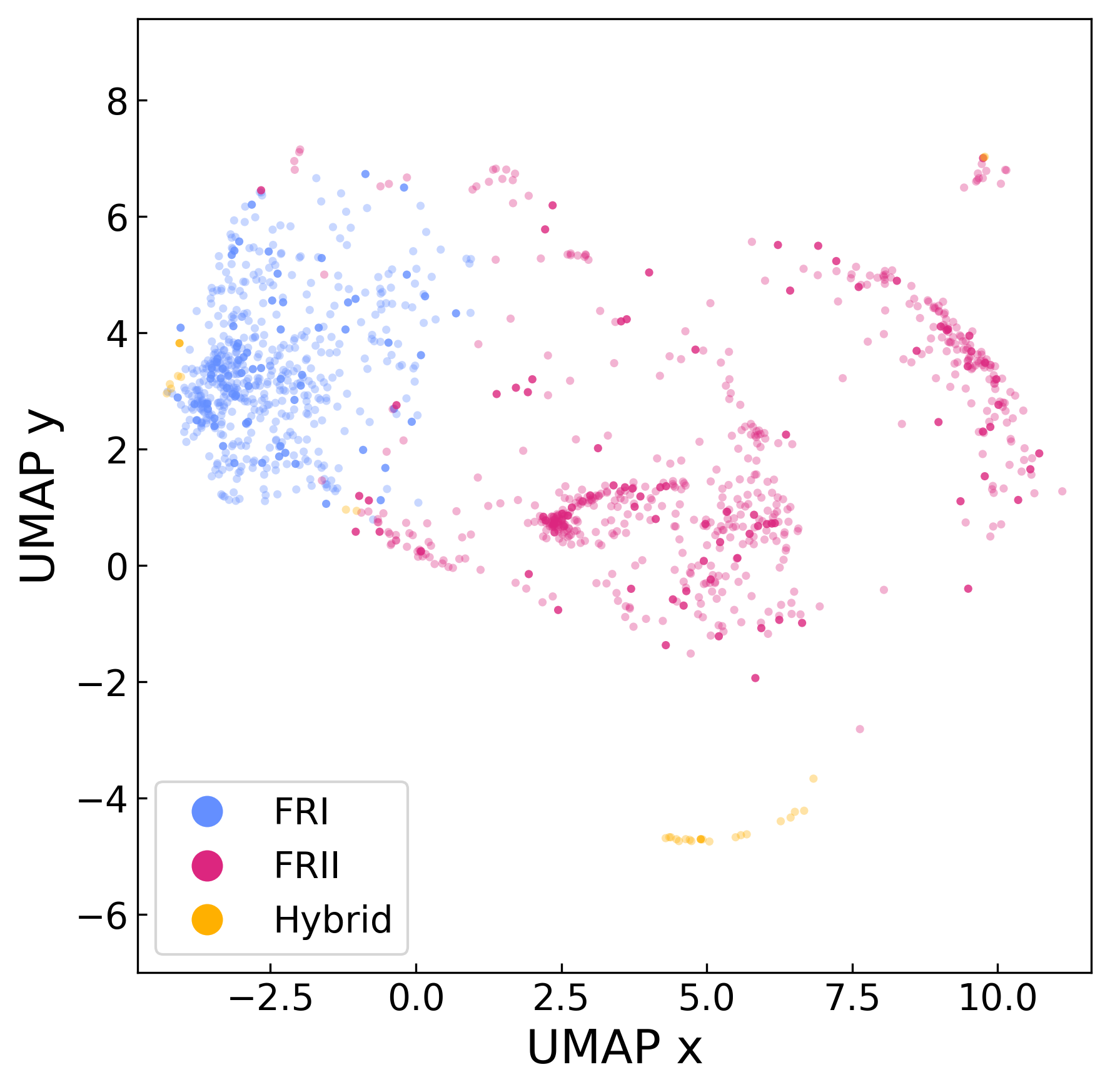}
    \centerline{(b)}
    \caption{Embedding of MiraBest data points from the model finetuned on the original MiraBest labels, colour-coded by (a) majority-vote annotator label and (b) model prediction. Note that the x and y dimensions here have no physical meaning.}
    \label{fig:embedding-2}
\end{figure}

To investigate whether the size of the training set and the use of majority-voted classes as labels introduced additional label noise, we compared these results to those from a BYOL pre-trained model that was fine-tuned on the original MiraBest labels, without the calibration set withheld. This model achieved higher accuracy (91\% on the uncertain test set and 98\% on the confident test set). The embedding is shown in Figure~\ref{fig:embedding-2}, and indicates that this model achieved slightly better class separation, particularly for the hybrid class. However, while the majority-voted class labels produce an embedding with a tightly clustered FRI class and quite sparsely distributed FRII class, the opposite appears to be true for the original MiraBest labels.

\subsection{Uncertainty quantification}
\label{sec:uq_results}
\begin{figure} 
\centering
    \includegraphics[width=\linewidth,trim={0cm 0cm 0cm 0cm},clip]{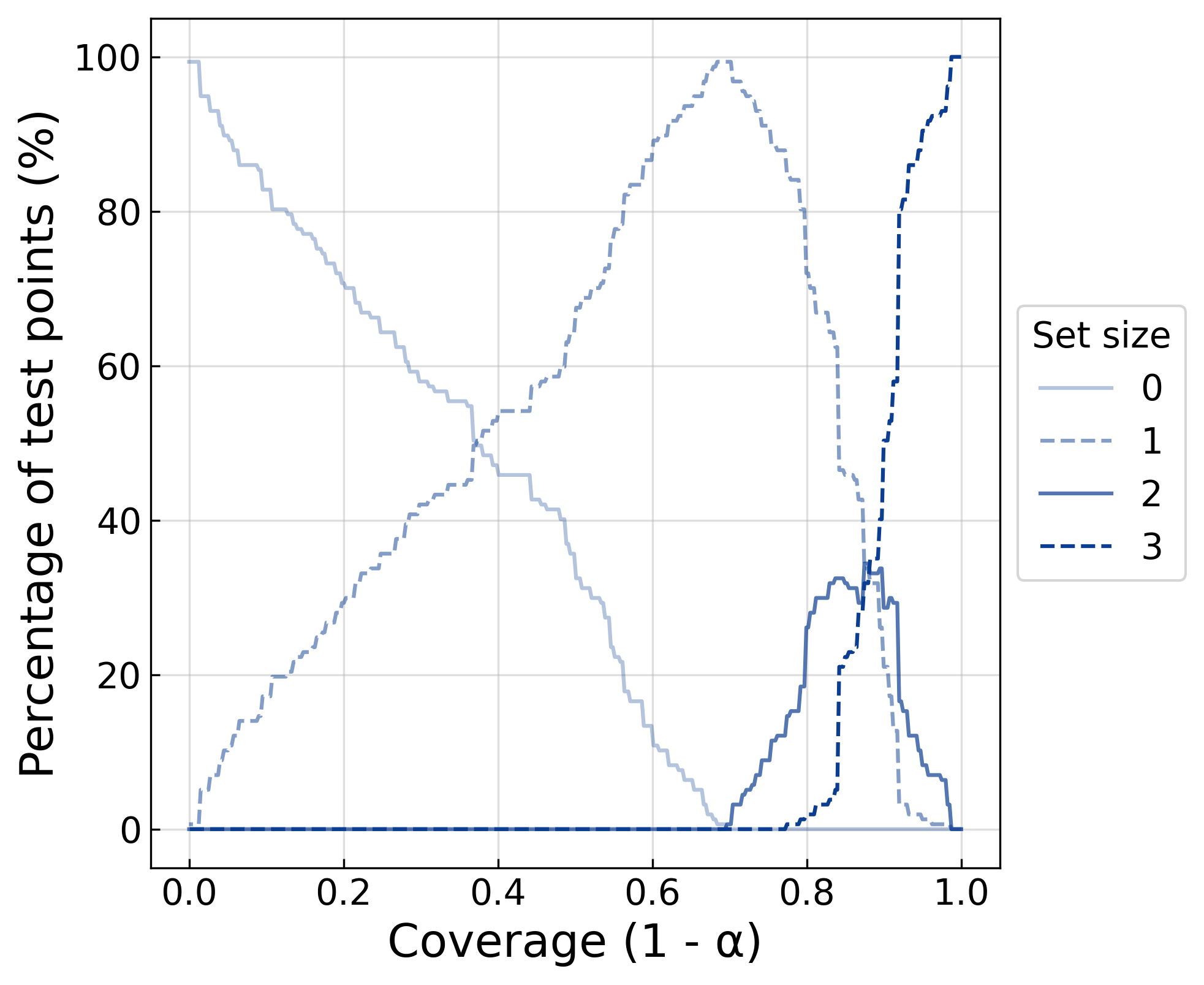}
    \caption{The percentage of test set samples with each possible prediction set size as a function of coverage, $1 - \alpha$, for $m$ = 100. At 100\% coverage ($1 - \alpha$ = 1), the prediction set must contain the correct class, so all prediction sets contain all of the possible classes.}
    \label{fig:uncertainty-1}
\end{figure}

Once the model was trained, $m$ = 100 labels were sampled from each point in the withheld calibration set, and non-conformity scores for the sampled labels were obtained using Equation \ref{eq:non-conformity}. Values of coverage between 0\% and 100\% were trialled to find the best trade-off between coverage level and the distribution of prediction set sizes. Figure \ref{fig:uncertainty-1} shows how the percentage of test samples with each possible prediction set size changed with coverage level.

\begin{figure}
    \includegraphics[width=\linewidth,trim={0cm 0.3cm 0cm 0.2cm},clip]{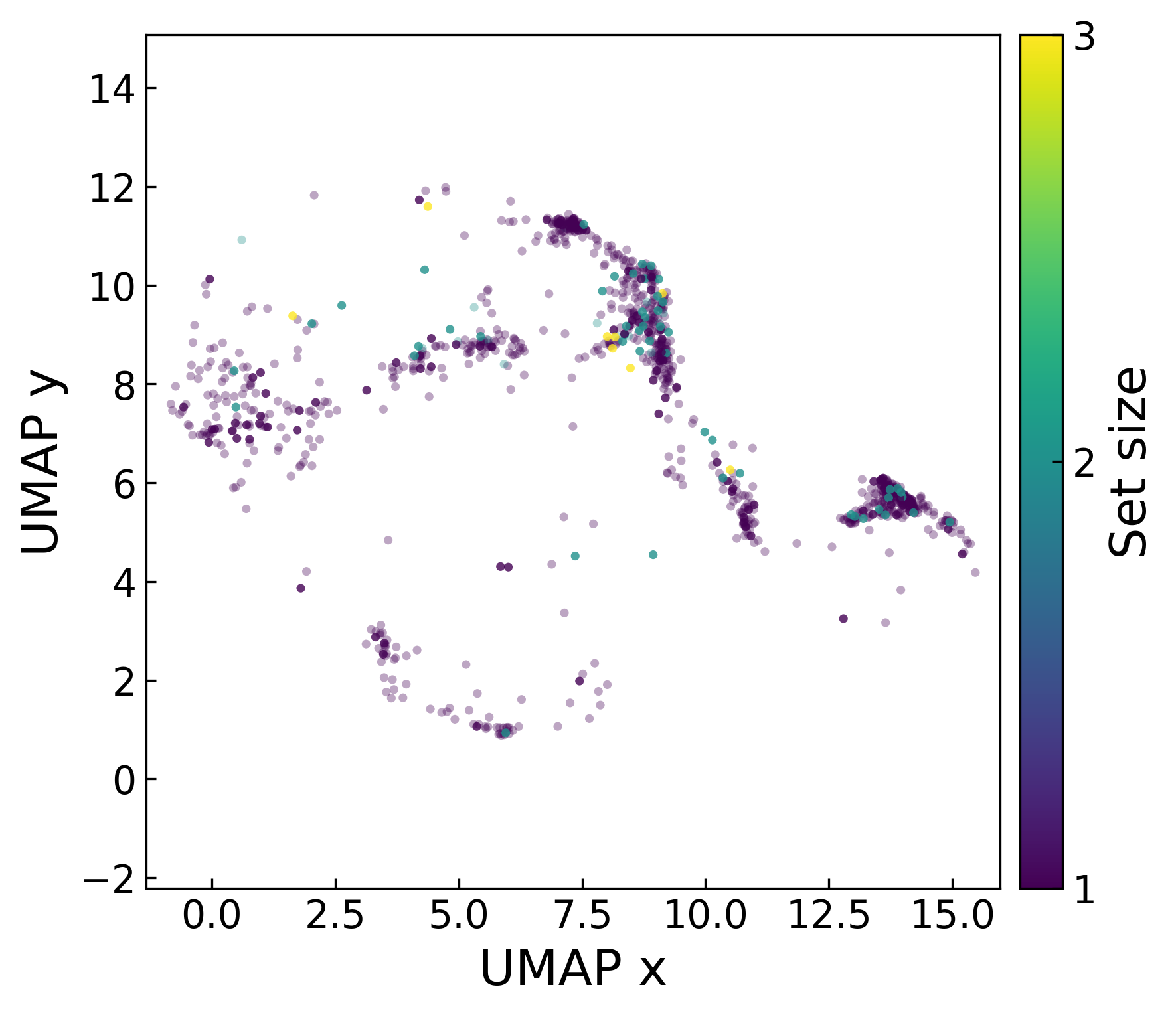}
    \centerline{(a)}
    \includegraphics[width=\linewidth,trim={0cm 0.3cm 0cm 0.2cm},clip]{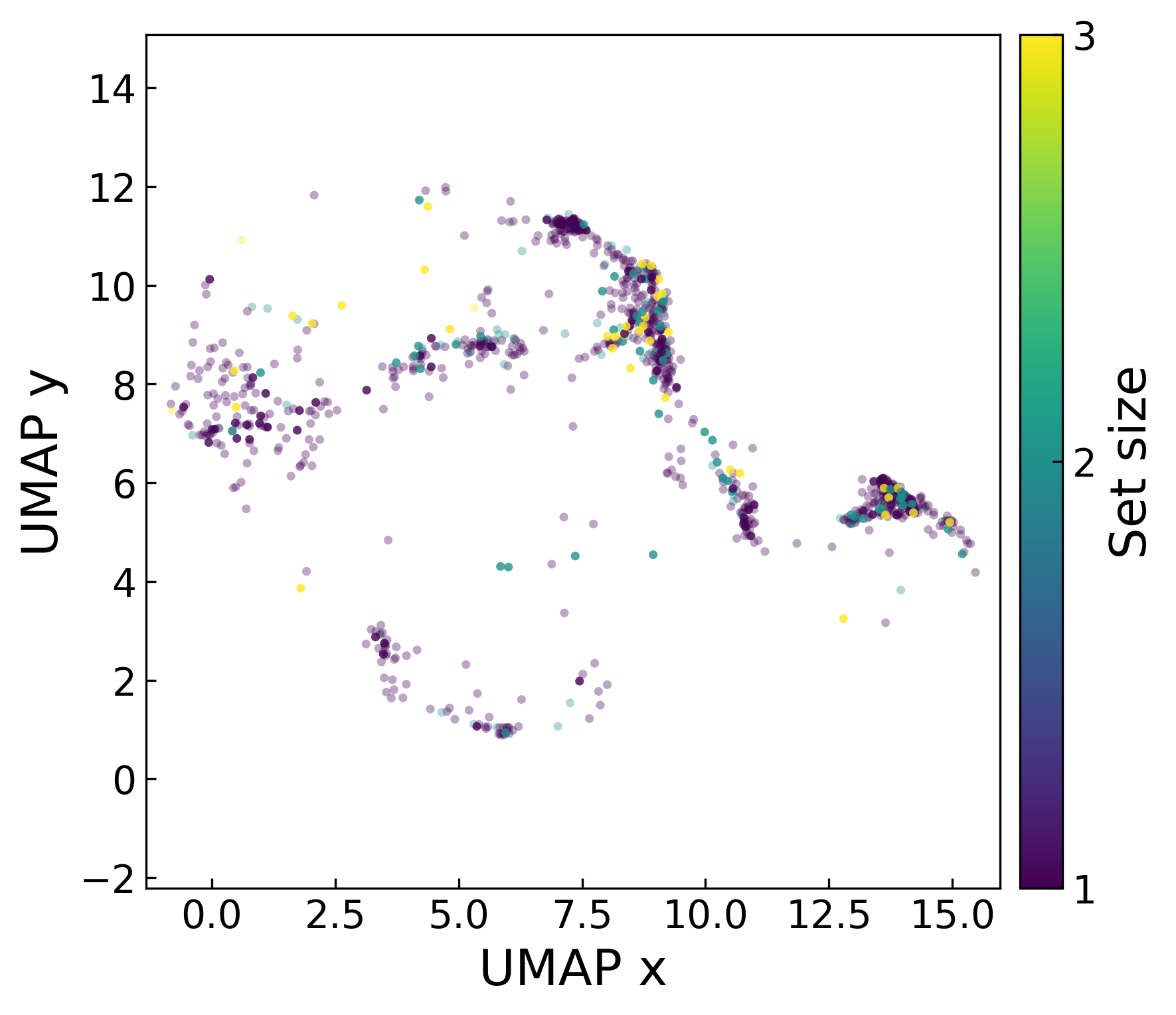}
    \centerline{(b)}
    \caption{Embedding of MiraBest data points colour-coded by (a) prediction set size for 84\% coverage and (b) prediction set size for 85\% coverage. Points that were included in the training set are plotted with lower opacity. Note that the x and y dimensions here have no physical meaning. 
    }
    \label{fig:uncertainty-2}
\end{figure}

As coverage increased above 80\%, single class prediction sets dropped off sharply, with two and three class prediction sets rapidly taking over at approximately the same level of coverage. This meant that an increase in coverage of only 1\% could produce a dramatic difference in the distribution of prediction set sizes. Figure \ref{fig:uncertainty-2} demonstrates this by showing the difference between 84\% coverage, where almost all prediction sets contain a single class, and 85\% coverage, where a significant proportion of prediction sets increase to two classes or three classes. Notably, some points changed directly from single class prediction sets to three class prediction sets, further demonstrating the high sensitivity to small changes in coverage.
This sensitivity indicates that the distribution of non-conformity scores is not smoothly decreasing towards 1 due to the limited size of the dataset, and we confirm that this is consistent with the behaviour of the calibration set used here.

This sensitivity makes determining an optimal value of coverage challenging. The sudden jumps in prediction set sizes, as well as the fact that almost all prediction sets contain a single label before any prediction sets expand to two labels, made Equation~\ref{eq:average-coverage} a poor indicator of the optimal coverage level. Instead, we chose the highest coverage level where the number of test samples changed monotonically with prediction set size, which was 87\%. From the distribution in Figure~\ref{eq:empirical-coverage-distribution}, the 95\% uncertainty interval for the empirical coverage at this level was [80\%, 94\%].

\begin{figure}
    \includegraphics[width=\linewidth,trim={0cm 0.3cm 0cm 0.1cm},clip]{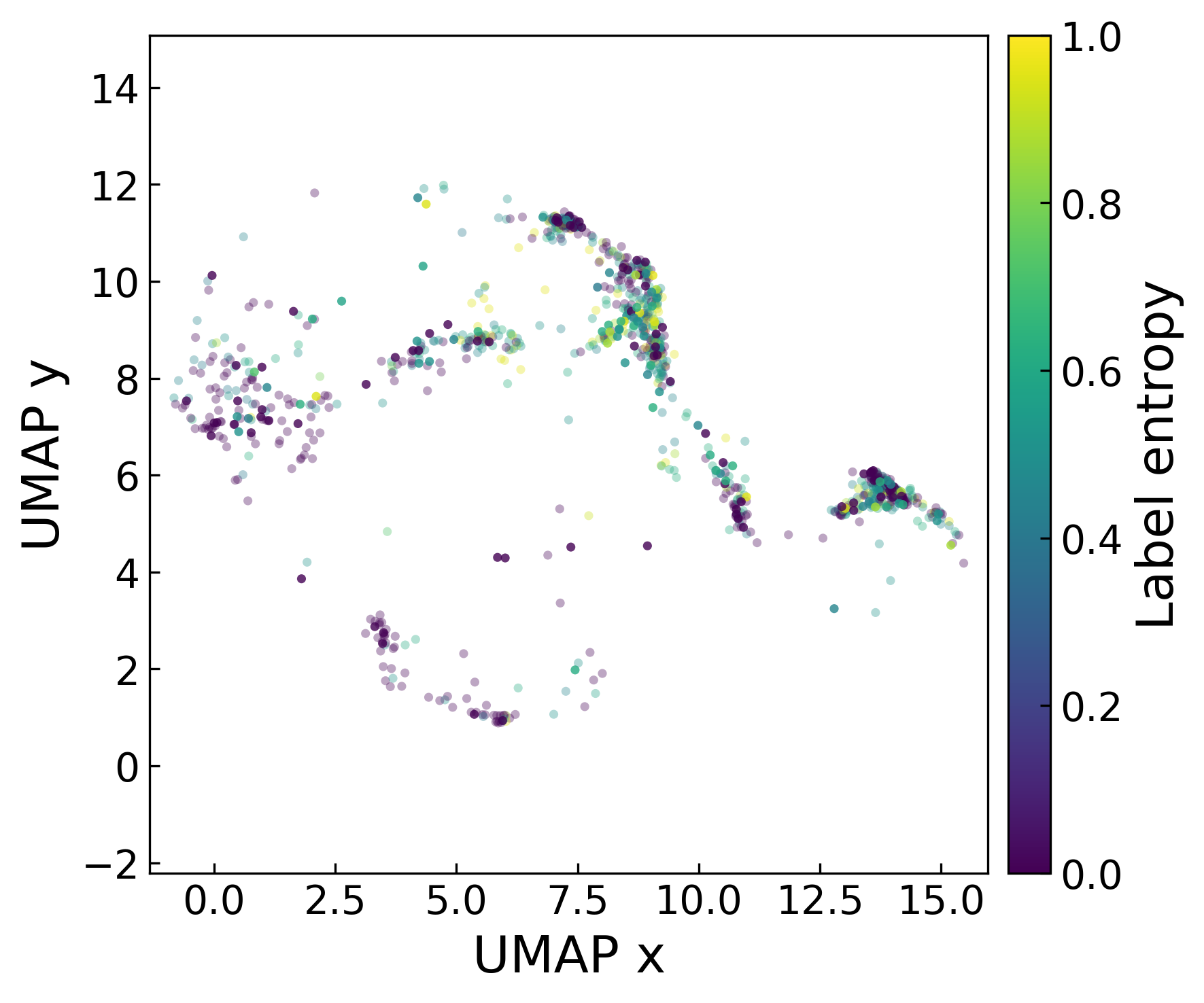}
    \centerline{(a)}
    \includegraphics[width=\linewidth,trim={0cm 0.3cm 0cm 0.1cm},clip]{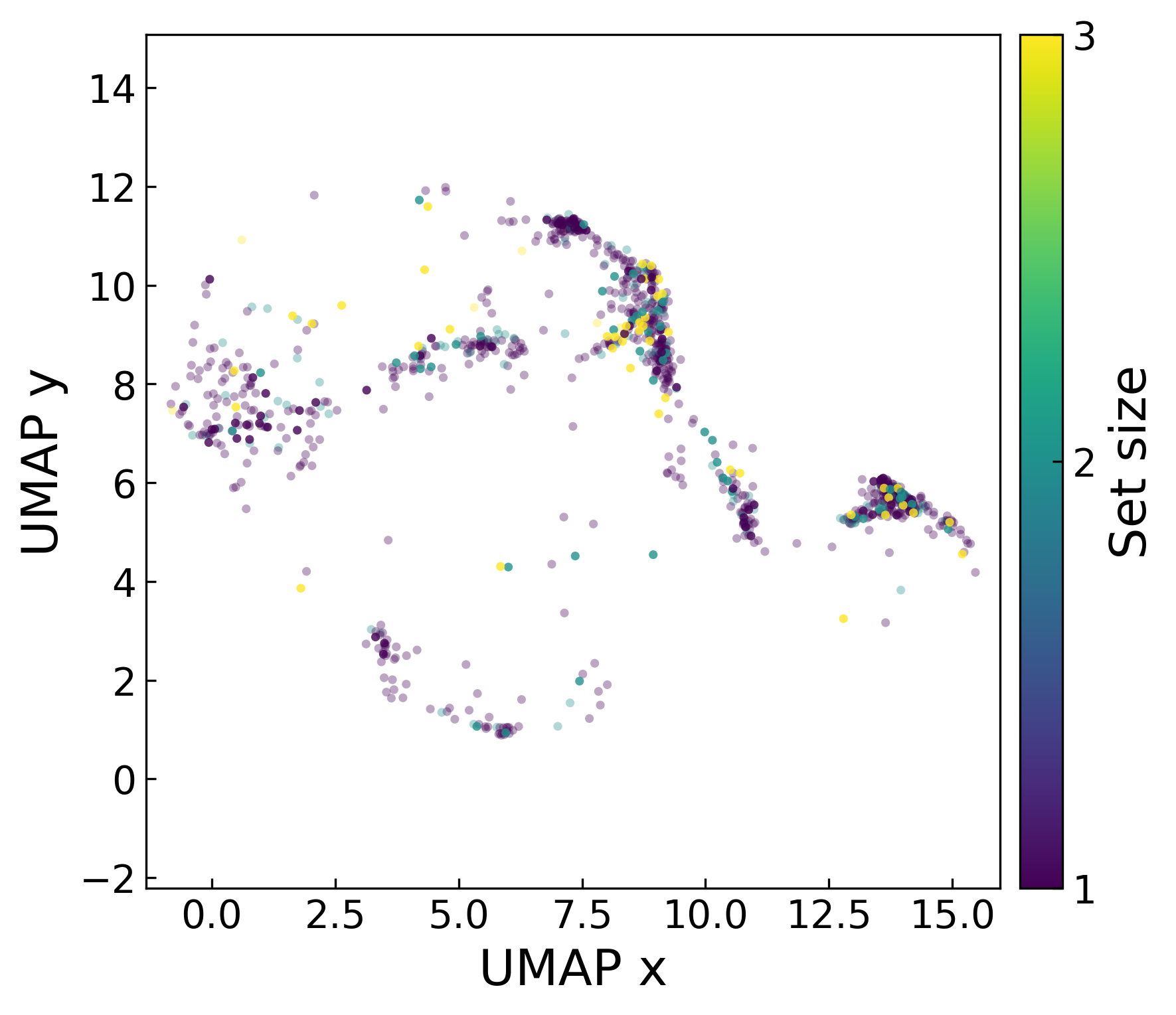}
    \centerline{(b)}
    \includegraphics[width=\linewidth,trim={0cm 0.3cm 0cm 0.1cm},clip]{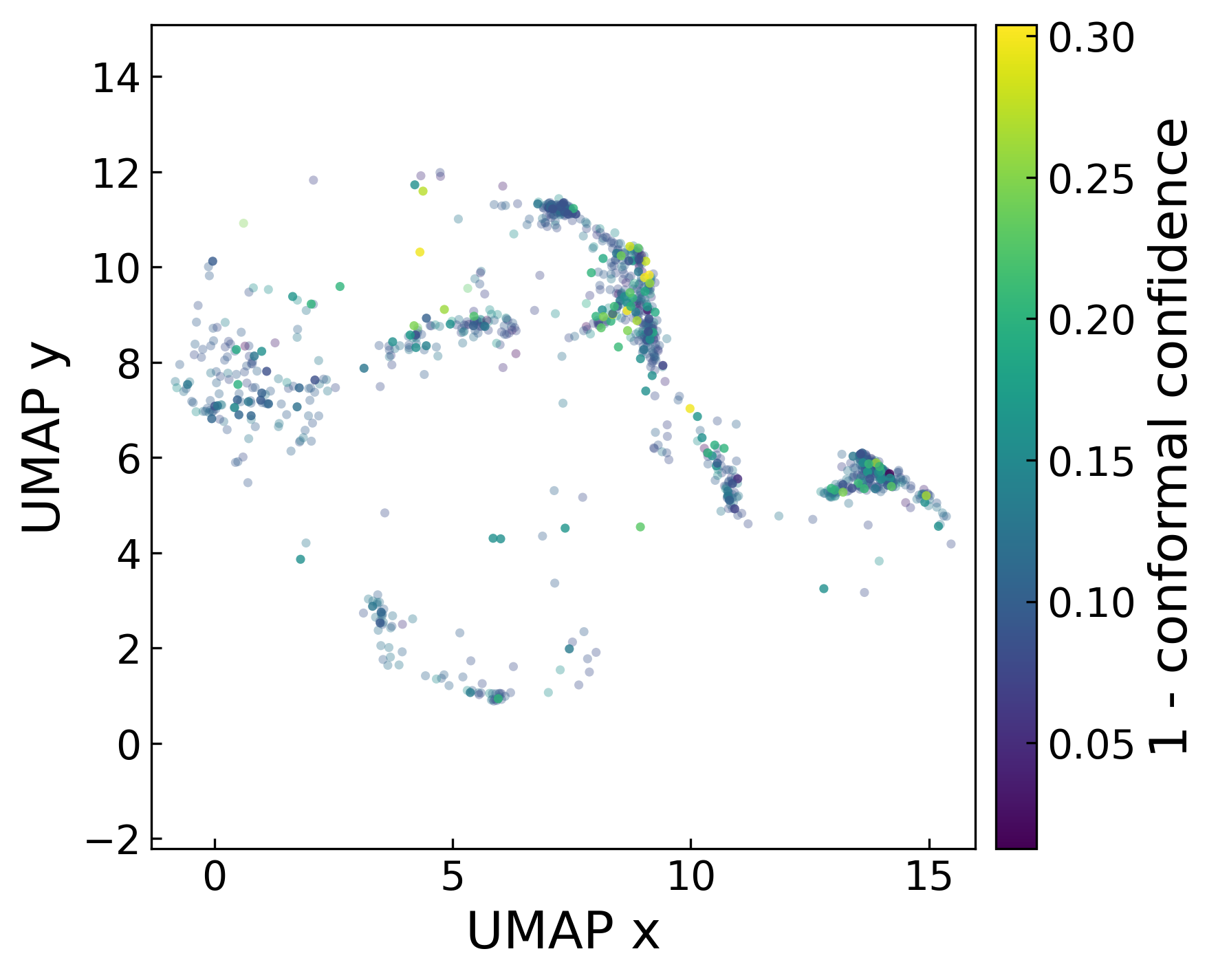}
    \centerline{(c)}
    \caption{Embedding of MiraBest data points colour-coded by (a) entropy of the annotator-derived label distribution, (b) prediction set size for 87\% coverage, and (c) 1 - conformal confidence. Points that were included in the training set are plotted with lower opacity. Note that the x and y dimensions here have no physical meaning.}
    \label{fig:uncertainty-3}
\end{figure}

Figure \ref{fig:uncertainty-3} compares the entropy of the annotator-derived soft label distribution with the prediction set sizes at 87\% coverage and the complement of the conformal confidence. In general, annotators were most uncertain about data points lying close to boundaries between classes, as can be seen by comparison with Figure \ref{fig:embedding-1}. The prediction set sizes and conformal confidence also reflect this trend. Annotators were also more uncertain about FRI sources than FRII sources; the average entropy over the label distributions of voted FRI sources was 0.40, while the same measure for voted FRII sources was 0.25. This is consistent with how the annotators reported their approach to the Zooniverse survey. It is important to note that low granularity in both the prediction set sizes and the entropy of the annotator distribution make comparison difficult. For this reason, it is only possible to observe broad trends.

\begin{figure}
\includegraphics[width=\linewidth]{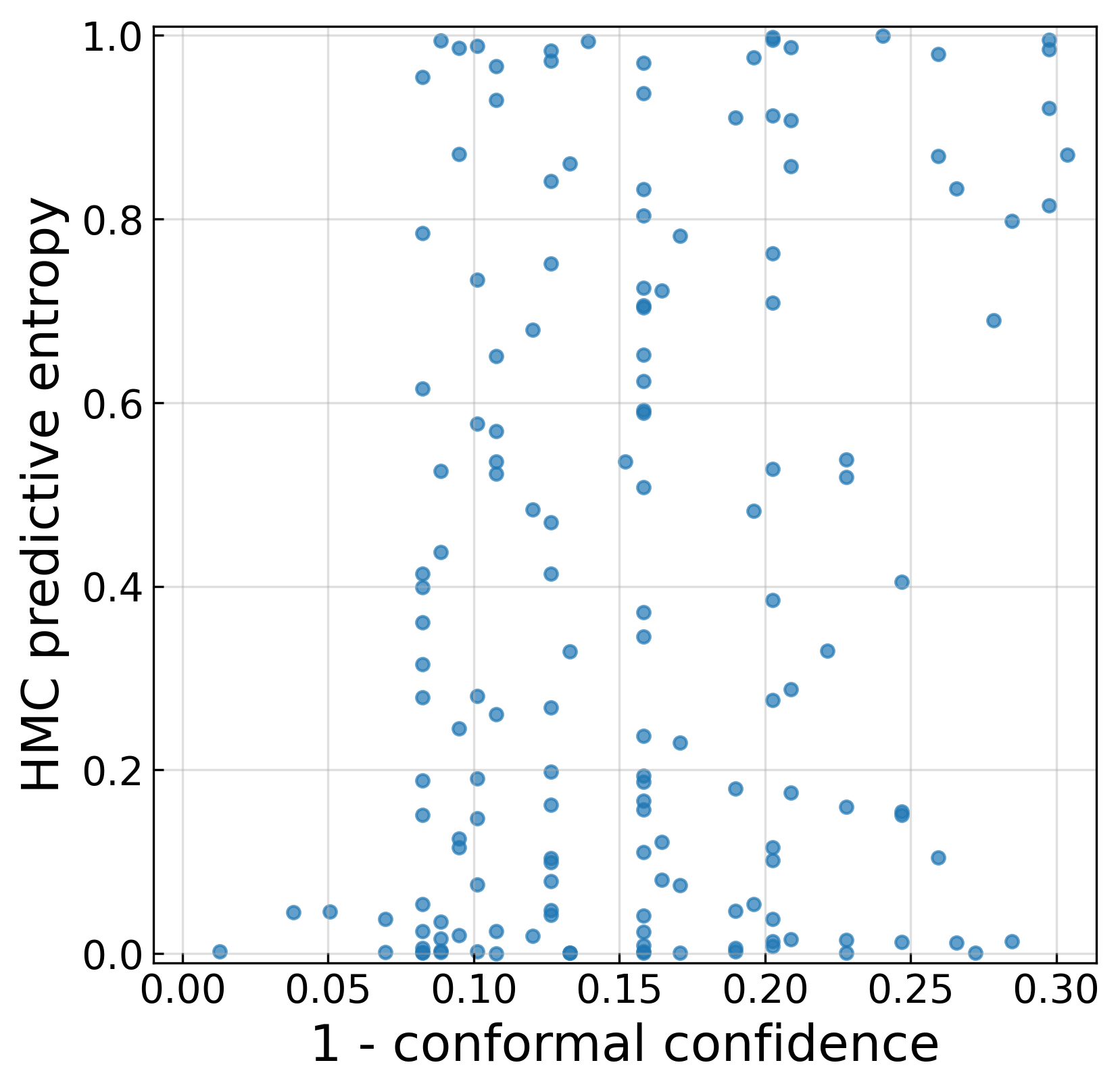}
    \caption{Predictive entropy values for images from the MiraBest test set obtained using the BNN trained by \citet{BNN} as a function of (1 - conformal confidence) from the MCCP analysis in this work.}
    \label{fig:uncertainty-4}
\end{figure}

We compared the prediction set sizes and conformal confidence scores obtained by our model to the values of predictive entropy obtained by the Hamiltonian Monte Carlo based BNN trained by \cite{BNN}. As the BNN was also trained on the MiraBest dataset, it was possible to directly compare how the two uncertainty measures performed on the same images. A comparison of the predictive entropy values and the complement of the conformal confidence scores for the MiraBest test set is shown in Figure \ref{fig:uncertainty-4}. Predictive entropy and prediction set size have a weaker positive correlation (Pearson correlation co-efficient $p=0.15$) than predictive entropy and the complement of conformal confidence ($p=0.21$). Both measures had a weaker correlation relative to the equivalent correlation of the entropy of the annotator-derived label distribution with the prediction set size ($p=0.30$) and complement of conformal confidence ($p=0.26$). 


\section{Discussion}
\label{sec:discussion}

\subsection{Impact of human annotators}
\label{human-behaviour}

As mentioned at the end of Section \ref{zooniverse}, annotators of the Zooniverse survey tended to label a galaxy as FRII if they were uncertain. This behaviour may have introduced several biases into the dataset. By systematically labelling ambiguous or low-signal sources as FRII, it is likely that the proportion of FRII-labelled sources in the dataset was artificially inflated. Furthermore, systematic labelling of faint or noisy sources gave them low entropy, often unanimous label distributions that did not reflect the uncertainty of the annotators. MCCP relies on sampling labels from a label distribution that reflects the ambiguity of the data. If a calibration image's label distribution indicates low ambiguity when the actual ambiguity is high, its non-conformity score will be artificially low. Across the entire calibration set, this has the effect of artificially lowering the threshold score. This results in overly narrow prediction sets and violates the coverage guarantee \citep{monte-carlo-cp}. This means that annotator behaviour may have caused under-coverage despite the use of MCCP. However, as we do not have access to the ground truth labels in this case, it is not possible to verify this.

Additionally, since the entropy on the annotator label distributions for ambiguous sources was compressed by annotator behaviour, there is almost no variation in the non-conformity scores calculated for ambiguous sources. This causes a spike in almost-identical non-conformity scores, which explains the sudden sharp decline in single class prediction sets and subsequent rapid rise in two and three class prediction sets for a small increase in coverage, as seen in Figure~\ref{fig:uncertainty-1} and Figure~\ref{fig:uncertainty-2}. The conformal confidence is also affected; since prediction sets do not begin to expand to contain two labels until coverage rises above approximately 70\%, the range of confidence scores is limited.

This effect was possibly exaggerated by a low number of annotations. Although 4.7 annotations per image is consistent with previous work using MCCP for uncertainty quantification, it may be too low to capture the range of annotator uncertainty in our dataset. Previous work has predominantly focused on medical imaging applications, where the ground truth can later be revealed by biopsy or exploratory surgery, allowing refinement of subsequent annotation protocols which reduces disagreement between expert annotators \citep{monte-carlo-cp, aortic-stenosis}. For radio galaxies, obtaining the ground truth morphology is not possible, so it is more difficult to refine the annotation process and more extensive labelling is needed to ensure disagreement between annotators is captured.

Figure \ref{fig:embedding-1} and Figure \ref{fig:embedding-2} reveal that embedding geometry is also affected by the behaviour of human annotators. When trained on majority-vote annotations, the model produces a sharply compact FRI cluster and a much less well-defined grouping of FRIIs. In contrast, the model trained on MiraBest labels produces a tight FRII cluster and a more scattered FRI distribution. While the MiraBest-trained model also benefits from a larger training split (as the calibration set was not withheld), the work by \cite{RGZ-BYOL}, which used the same architecture on a much smaller training data split and achieved comparable accuracy, suggests that label bias is the primary driver of this effect. Because annotators defaulted to FRII when uncertain, only the most unambiguous FRI sources were given the FRI label, explaining its compact embedding. Meanwhile, it is likely that FRII-labelled sources were disproportionately misclassified, which hindered the model's ability to learn a coherent FRII representation.
The clustering behaviour also suggests that the finetuned embeddings might be robust to biases in the finetuning data. The prior on FRII-like galaxies from the pretrained foundation is strong enough that the biased FRII human annotations in the finetuning data do not 
lead to artificially tight clustering one would expect due to annotator bias.

Due to the systematic responses of annotators to ambiguous morphologies, soft label distributions derived from a normalised set of annotations do not reflect the actual ambiguity in the data, which compromises the statistical validity of MCCP as a method of uncertainty quantification and introduces biases that influence the representation learned by the model. In order to use MCCP for radio galaxy classification, an alternative method for deriving a soft label distribution is required. This could be achieved by adopting the approach detailed in work on modelling annotator certainty by \cite{annotator-ambiguity-modelling}, in which annotations are combined with a separate, self-reported measure of confidence used to weight the derived label distribution. This approach would ensure that even if most annotators select FRII in ambiguous cases, low confidence scores would preserve high entropy in the soft label distributions. Acquiring more annotations per image would also help to ensure that soft label distributions accurately reflect ambiguity. However, collecting enough annotations across a sufficiently diverse image set to capture all morphological variations is a significant challenge, highlighted by the fact that it is the primary motivation for turning to automated classification methods in the first place.

\subsection{Comparison of MCCP and BNNs for uncertainty quantification}
\label{BNN-comparison}

As shown by Figure \ref{fig:uncertainty-4}, MCCP conformal confidence and predictive entropy from the BNN trained by \cite{BNN} appear to share little to no agreement.
Predictive entropy is a measure of the sharpness of the posterior predictive distribution (how concentrated the BNN's predictions are on a single class). While it is a combined measure of both aleatoric and epistemic uncertainty, there is no mechanism by which it can take into account inherent label ambiguity caused by human uncertainty \citep{PE-mmislead}. This is further supported by the observation of \cite{BNN-uncertainty} that their BNN can assign low predictive entropy scores (indicating high confidence) to misclassified sources with ambiguous or potentially out-of-distribution morphology. In contrast, the coverage guarantee enforced by MCCP ensures that prediction set size reflects inherent label ambiguity \citep{monte-carlo-cp}. \cite{cp-human-uncertainty} suggest that uncertainty measures derived from MCCP mimic human uncertainty, and the broad agreement between the entropy of the soft label distributions, predictions set sizes and conformal confidence scores (particularly for the most uncertain points) shown in Figure~\ref{fig:uncertainty-3} supports this. This fundamental difference in the type of uncertainty each measure represents explains
why the agreement between them is so poor.


As noted in Section \ref{sec:pred_set_size}, the MCCP primarily captures aleatoric uncertainty due to label ambiguity. To improve MCCP, there is a need to incorporate measures of epistemic uncertainty in the outputs. 
Recent work in this direction  includes Epistemic Conformal Score (EPICSCORE) which captures epistemic uncertainty in  first-order conformal scores using various Bayesian methods such as Gaussian Processes, Bayesian additive regression trees, and neural networks with Monte Carlo (MC) dropout \citep{cabezas2025epistemic}. An alternative approach involves credal sets, which are convex sets of plausible first-order probabilities. \citet{caprio2025conformalized} construct conformalised credal regions using plausibility regions and extend their approach to classification problems under ambiguous ground truth and  
\citet{javanmardi2025optimal} propose Bernoulli Prediction Sets  to construct the smallest possible prediction sets with conditional coverage for any distribution within the credal set using various second order predictors such as deep ensembles, MC dropout, and conformalised set predictors. These methods will be explored in future work.

\section{Conclusions}
\label{sec:conclusions}


In this work we have created a framework for using Monte Carlo conformal prediction (MCCP) to return measures of uncertainty on radio galaxy classifications, which are inherently ambiguous in terms of their ground truth. We have evaluated this approach in comparison to an alternative measure of uncertainty from predictive entropy using Bayesian neural networks (BNN). Our analysis reveals limited agreement between these measures. High-entropy cases tend to coincide with larger conformal sets, but low-entropy examples appear almost uniformly across all set sizes. This divergence supports the idea that each method provides a fundamentally different picture of the overall uncertainty: entropy reflects the sharpness of the model’s internal posterior, whereas MCCP enforces an empirical coverage level that automatically captures inherent label ambiguity. However, we also note that in the case of finite (limited) datasets where subjective human annotations are used for data labelling, these additional complexities will provide additional confounders that limit such comparisons.

Our results also highlight that the behaviour of human annotators can compromise the coverage guarantee of MCCP and hinder a model's ability to learn a coherent representation of the data. Refining annotation protocols to produce soft label distributions that reflect the true level of ambiguity in the data will be necessary for MCCP to be used reliably for astronomical applications in the future. Additionally, our results have also demonstrated the sensitivity of the conformity score distribution to model performance, highlighting the effect that  non-smooth conformity score distributions can have on predicted set sizes, as illustrated in Section~\ref{sec:uq_results}. This effect is also potentially influenced by annotator behaviour, which is incorporated into the creation of the conformity score distribution through label realisations drawn from the annotator label distribution of the calibration set as part of the MCCP approach. Consequently, we conclude that the soft-label distributions required by the calibration set for MCCP are potentially problematic in multiple ways in the context of astronomical classification under ambiguous ground truth.

\section*{Acknowledgements}

AW and DM gratefully acknowledge support from the University of Manchester. AMS gratefully acknowledges support from the UK Alan Turing Institute under grant reference EP/V030302/1.

The authors would like to thank Emma Alexander, Philip Best, Bernie Fanaroff, Katie Hesterly, Neal Jackson, Larry Rudnick, and Hongming Tang for providing both annotations and useful discussions that contributed to this project. 

The authors would like to thank the reviewers for their comments, which improved the content of this paper.

This work has been made possible by the participation of more than 12,000 volunteers in the Radio Galaxy Zoo Project. The RGZ data in this paper are the result of the efforts of the Radio Galaxy Zoo volunteers, without whom none of this work would be possible. Their efforts are individually acknowledged at \href{http://rgzauthors.galaxyzoo.org}{http://rgzauthors.galaxyzoo.org}.

\section*{Data Availability}


The self-supervised learning approach used in this work was previously published at \cite{RGZ-BYOL}, and the code for the model is available at \url{https://github.com/inigoval/}. Standardised catalogues of labelled radio source images, including MiraBest, which were used to train the model, are accessible at   ~\url{https://doi.org/10.5281/zenodo.4288837} and \url{https://doi.org/10.5281/zenodo.8188867}, respectively. 

Code for reproducing the results in this work, including all figures from the paper, can be found at \url{https://github.com/alexwalls01/byol}.



\bibliographystyle{mnras}
\bibliography{mccp_paper} 







\bsp	
\label{lastpage}
\end{document}